\documentclass[aps,superscriptaddress,showpacs,nofootinbib, 10pt]{revtex4-1}
\usepackage{mathrsfs}
\usepackage{bigints}
\usepackage{latexsym,bm}
\usepackage{graphicx}
\usepackage{indentfirst}
\usepackage{slashed}
\usepackage{amssymb}
\usepackage{amsmath}
\usepackage{bbm}
\usepackage{color}
\usepackage[dvipsnames]{xcolor}
\usepackage{epsfig}
\usepackage[titletoc]{appendix}
\usepackage{multirow}%
\usepackage{rotating}
\usepackage{epstopdf}
\usepackage{extarrows}
\usepackage{tabularx}
\usepackage{soul} 
\usepackage{xcolor}
\usepackage{lipsum}
\usepackage[colorlinks=true,allcolors=BlueViolet,hyperfootnotes=false]{hyperref}

\usepackage[utf8]{inputenc}

\newcommand{\be}{\begin{equation}} \newcommand{\ee}{\end{equation}}
\newcommand{\ba}{\begin{array}{c}} \newcommand{\ea}{\end{array}}
\newcommand{\bea}{\begin{eqnarray}} \newcommand{\eea}{\end{eqnarray}}
\newcommand{\beas}{\begin{eqnarray*}} \newcommand{\eeas}{\end{eqnarray*}}

\newcommand{\no}{\nonumber\\}

\begin{document}
\title{ Probing New Physics and CP Violation in $\nu_\tau n \to \Lambda_c  \tau^- (\pi^- \nu_\tau)$ and $\bar\nu_\tau p \to \Lambda  \tau^+ (\pi^+ \bar\nu_\tau)$}

\author{E. Hern\'andez}
\affiliation{Departamento de F\'\i sica Fundamental 
  e IUFFyM,\\ Universidad de Salamanca, E-37008 Salamanca, Spain}

\author{J. Nieves}
\affiliation{Instituto de F\'{\i}sica Corpuscular (centro mixto CSIC-UV), 
Institutos de Investigaci\'on de Paterna,
Apartado 22085, 46071, Valencia, Spain}

\author{J. E. Sobczyk}
\affiliation{Department of Physics, Chalmers University of Technology, SE-412 96 G\"oteborg, Sweden}

\date{\today}

\begin{abstract}
We study the processes $\nu_\tau n \to \Lambda_c \tau^- (\pi^- \nu_\tau)$ and $\bar\nu_\tau p \to \Lambda \tau^+ (\pi^+ \bar\nu_\tau)$, with particular emphasis on the pion energy and angular distributions, as a possible signal for lepton flavor universality violation, and in general of physics beyond the Standard Model (SM), as well as a sensitive probe of the $\tau$ polarization vector.
We work within an effective low-energy extension of the SM with all dimension-six four-fermion operators. In this framework, complex Wilson coefficients which encode new physics can generate CP-violating contributions. We propose an observable that provides a genuine CP-odd signal due to its sensitivity to particular transverse  components of the $\tau$ polarization vector. Namely, we show that the asymmetry in the azimuthal-angle distribution of the pion in the decay $\tau^\pm\to \pi^\pm \nu_\tau$ constitutes a smoking-gun prediction of such a beyond the SM scenario. We estimate the strength of this effect extrapolating nucleon-hyperon form factors recently obtained from lattice QCD calculations.

\end{abstract}

\maketitle

\section{Introduction}

One of the central features of the Standard Model (SM) is lepton flavor universality (LFU), which states that the three lepton families couple identically to the $W$ and $Z$ gauge bosons. However, several experimental measurements have suggested the possible presence of LFU violation. The most compelling indications of possible physics beyond the SM (BSM) arise from the experimentally averaged values of the ratios ${\cal R}_{D^{(*)}}=\Gamma(\bar B\to D^{(*)}\tau^-\bar\nu_\tau)/
\Gamma(\bar B\to D^{(*)}\ell^-\bar\nu_\ell)$, which display a combined discrepancy with SM predictions at the level of $3.1\sigma$, as reported by the HFLAV Collaboration~\cite{HeavyFlavorAveragingGroupHFLAV:2024ctg}. This tension suggests the presence of non-universal charged-current interactions affecting the $\tau$ lepton. Further support for this interpretation is provided by the ratio ${\cal R}_{J/\psi}=  \Gamma(\bar B_c\to J/\psi\tau^-
\bar\nu_\tau)/\Gamma(\bar B_c\to    J/\psi\mu^-\bar\nu_\mu)$ measured by the LHCb Collaboration~\cite{LHCb:2017vlu}, which also exhibits an approximate $1.8\sigma$ deviation from SM expectations~\cite{Anisimov:1998uk,Ivanov:2006ni,
Hernandez:2006gt,Huang:2007kb,Wang:2008xt,Wen-Fei:2013uea, Watanabe:2017mip, Issadykov:2018myx,
Tran:2018kuv, Hu:2019qcn,Leljak:2019eyw,Azizi:2019aaf,Wang:2018duy}, reinforcing the case for potential new physics (NP) in semileptonic $b\to c\tau\nu$ transitions.

Other measurements, on the other hand, are fully consistent with LFU. This is the case for the ratio ${\cal R}_{\Lambda_c}=
  \Gamma(\Lambda^0_b\to \Lambda^+_c\tau^-\bar\nu_\tau)/\Gamma(\Lambda^0_b\to 
  \Lambda^+_c
  \ell^-\bar\nu_\ell)$,  measured by the LHCb Collaboration~\cite{LHCb:2022piu}, which agrees with SM predictions based on the theoretical analysis of Ref.~\cite{Bernlochner:2018bfn}. A similar conclusion is reached from the recent first measurement of the inclusive semileptonic decay ratio ${\cal R}_{X_{\tau/\ell}}=\Gamma(\bar B\to X\tau^-\bar\nu_\tau)/\Gamma(\bar 
B\to X\ell^-\bar\nu_\ell)$,  reported by the Belle II Collaboration~\cite{Belle-II:2023aih}, which is also compatible with SM  expectations as evaluated in Refs.~\cite{Ligeti:2021six,Rahimi:2022vlv}. These results provide important complementary constraints on possible NP scenarios and highlight the need for further precise and diverse tests of LFU across different decay channels.

Whether the currently observed deviations from LFU will persist under increased experimental precision remains an open question. Should these anomalies be confirmed, physics BSM would be required to account for the data, and it would be natural to expect such effects to impact, to varying degrees, all quark and lepton generations. Given that the present experimental evidence is not yet conclusive, any NP contributions are expected to be moderate, at least at the energy scales probed by current experiments.

Effective field theory (EFT) offers a systematic and model-independent framework to parametrize the effects of physics BSM at energies well below the characteristic NP scale. In this formalism, the low energy SM Lagrangian is extended to a complete basis of  dimension-six four-fermion operators describing charged-current quark-level transitions of the form $q \to q' \ell \nu$. These operators encode the virtual effects of heavy degrees of freedom associated with NP at scales much larger than those directly probed by current experiments. Early seminal work in this context can be found in Ref.~\cite{Fajfer:2012vx}. The impact of these operators on low-energy observables is controlled by the corresponding Wilson coefficients, which are, in general, complex and scale-dependent, and whose magnitudes and phases can be extracted or constrained through global analyses of precision semileptonic and leptonic  decay data.

In the case of the quark-level $d \leftrightarrow c$ transition, valuable information can be obtained from the leptonic decay-width ratio ${\cal  R}_{\tau/\mu}=\Gamma(D^+\to\tau^+\nu_\tau)/ \Gamma(D^+\to\mu^+\nu_\mu)$. This observable was first measured by the CLEO Collaboration~\cite{CLEO:2006jxt}, which reported a statistically non-significant result and provided only an upper bound. A subsequent measurement by the BESIII Collaboration~\cite{BESIII:2019vhn} yielded a central value larger than the SM expectation, although still consistent with it at the $1\sigma$ level. Within the EFT framework, this result has been extensively used to constrain the Wilson coefficients associated with charged-current $c \to d$ transitions, as discussed in Refs.~\cite{Fleischer:2019wlx,Becirevic:2020rzi,Leng:2020fei,Fuentes-Martin:2020lea}.
While the above purely leptonic decays\footnote{A final-state $\tau$ is kinematically 
forbidden due to phase-space.} are only sensitive to axial-vector and 
pseudoscalar Wilson coefficients, one has that vector, scalar, and tensor operators can be 
probed through  mono-lepton production in high-energy charged-current Drell–Yan reactions~\cite{Fuentes-Martin:2020lea}.

In Ref.~\cite{Kong:2023kkd}, the authors proposed the process $\nu_\tau n \to \Lambda_c\tau^-$ as a probe to further constrain the Wilson coefficients associated with the $d \leftrightarrow c$ quark transition. Their analysis was carried out at the nucleon level and did not include a dedicated treatment of nuclear effects. Since free neutron targets are not available, such effects are expected to be relevant—particularly for heavy nuclear targets—and may have a non-negligible impact on the phenomenology of this reaction.
In Ref.~\cite{Hernandez:2025snr}, we revisited this problem and derived general expressions for inclusive quasi-elastic (QE) reactions on nuclear targets, 
\begin{equation}
 (\bar \nu_\ell ) 
\nu_\ell A_Z \to (\ell^+)\ell^- Y X   
\end{equation}
where $Y$ denotes either a strange or a charmed hyperon, $\ell = e,\,\mu,\,\tau$, and $X$ represents the residual nuclear system with $(A-1)$ nucleons. In principle, such processes could be accessed at the Deep Underground Neutrino Experiment (DUNE) far detector, exploiting the oscillation of the initially produced $\nu_\mu/\bar\nu_\mu$ neutrinos and antineutrinos into $\nu_\tau/\bar\nu_\tau$. At  DUNE, a significant fraction of the neutrino flux is expected to lie above the $\tau$ production threshold of $E_{\nu_\tau}\gtrsim 3.5$ GeV. In the scenario considered here, the relevant thresholds are $E_{\bar\nu_\tau}\gtrsim 4$ GeV and $E_{\nu_\tau}\gtrsim 8.3$ GeV for $\Lambda$ and $\Lambda_c$ hyperons in the final state, respectively. In particular, $\Lambda_c$ production requires operation in the $\tau$-optimized beam mode, characterized by a broader energy spectrum (see, for instance, Fig.~1 of Ref.~\cite{Machado:2020yxl}). Within the three-neutrino paradigm, several studies have explored the potential of DUNE to constrain non-standard interactions affecting neutrino propagation in matter, as well as the possibility of direct $\nu_\tau$ production in pion decays. Both effects, if present, would modify the $\nu_\tau$ flux at the far detector.

In this work, we retake  our former calculation in Ref.~\cite{Hernandez:2025snr} and go a step further by analyzing the processes $\nu_\tau n \to \Lambda_c \tau^- (\pi^- \nu_\tau)$ and $\bar\nu_\tau p \to \Lambda \tau^+ (\pi^+ \nu_\tau)$, in which the final-state $\tau^\pm$ subsequently decays into the $\pi^\pm \nu_\tau$ channel. Incorporating the $\tau$ decay is essential for two main reasons. First, the $\tau$ is a short-lived particle and must be experimentally reconstructed through its decay products. Second, the kinematic and angular distributions of the final-state pion provide direct access to the full $\tau$ polarization vector, including its transverse components, which are sensitive to the presence of complex phases in the underlying Wilson coefficients. As a result, these observables offer a powerful handle on CP-violating effects and constitute a significantly richer probe of NP dynamics than analyses based solely on the reconstructed $\tau$ lepton.

At this stage, we refrain ourselves from including nuclear effects, since they obscure the treatment of a four-particle final state within a general BSM-EFT framework. Moreover, proton targets—or possibly deuteron targets—may become experimentally accessible in the future, further motivating the study of these reactions at the nucleon level. In a subsequent study, we will address these reactions with nuclear targets using a realistic many-body treatment.

Although the analyses of semileptonic $D$-meson decays and high-energy Drell–Yan processes performed in Ref.~\cite{Fuentes-Martin:2020lea} place stringent constraints on the Wilson coefficients, they do not exclude the possibility that these coefficients have complex values, which would imply the presence of CP-violating phases already at the Hamiltonian level.
In this work, we propose an observable that provides a genuine CP-odd and essentially smoking-gun signal of such a BSM scenario. Specifically, we advocate measuring the asymmetry in the azimuthal-angle distribution of the  pion in the decay $\tau^\pm \to \pi^\pm \nu_\tau$. Any imbalance between the number of charged pions emitted above and below the plane defined by the incoming neutrino and the outgoing hyperon constitutes a direct manifestation of CP violation and must be driven by NP. Importantly, this asymmetry is defined in terms of a ratio, which reduces its dependence on the hadronic form factors, a key advantage given the large theoretical uncertainties affecting their determination, particularly at high momentum transfers $Q^2=-q^2$.

This work is organized as follows. In Sec.~\ref{sec:eh}, we introduce the effective Hamiltonian and exploit the close correspondence between neutrino- and antineutrino-induced reactions, showing that, up to trivial modifications, the expressions derived for $\nu_\tau n \to \Lambda_c \tau^-$ also apply to $\bar\nu_\tau p \to \Lambda \tau^+$. In Sec.~\ref{sec:Lambdac}, we present the total cross section and the pion azimuthal-angle asymmetry for the process $\nu_\tau n \to \Lambda_c \tau^- (\pi^- \bar\nu_\tau)$, both of which depend on the $\tau$ polarization vector ${\cal P}$. We show that this asymmetry probes exclusively the transverse–transverse component ${\cal P}_{TT}$\footnote{Defined as perpendicular to the plane spanned by the incoming neutrino and the outgoing $\tau$ directions. Since the initial nucleon is taken at rest, this plane coincides with that defined by the  neutrino and the outgoing hyperon.} and becomes nonzero only in the presence of complex Wilson coefficients. In Sec.~\ref{sec:Lambda}, we describe the modifications required to treat the reaction $\bar\nu_\tau p \to \Lambda \tau^+ (\pi^+ \bar\nu_\tau)$ and specify the form-factor inputs used in this case. Our results are presented in section~\ref{sec:results}, and the main conclusions of this work are summarized in Section~\ref{sec:concl}. We also include two Appendices which provide some of the technical details involved in the analytical derivation of the cross sections.

 \section{Effective Hamiltonian}
 \label{sec:eh}
 For both the $\nu_\tau d \to c \tau^-$ and $\bar\nu_\tau u \to s\tau^+$ reactions, we consider the complete set of dimension-six semileptonic operators involving left- and right-handed neutrinos. Although these operators were already presented in Ref.~\cite{Hernandez:2025snr}, we reproduce them here for completeness and convenience.
 
For the $\nu_\tau d\to c\tau^-$ transition we have
\bea
 H^{\nu_\tau d\to c\tau^-}_{\rm eff}&=&\frac{4G_F V_{cd}}{\sqrt2}\Big\{\sum_{\chi,\chi'=L,R}\Big[
 (\bar c\gamma^\mu[\delta_{\chi'L}\delta_{\chi L}+C^V_{cd\tau\chi'\chi}]P_{\chi'}d)\ 
 (\bar\tau\gamma_\mu  P_\chi\nu_\tau)
 +C^S_{cd\tau\chi'\chi}\,(\bar c
 P_{\chi'}d)\ 
 (\bar\tau  P_\chi\nu_\tau)\Big]\no
 &&\hspace{3cm}+\sum_{\chi=L,R}C^T_{cd\tau\chi\chi}(\bar c\,\sigma^{\mu\nu}
 P_\chi d)\ 
 (\bar\tau \sigma_{\mu\nu} P_\chi\nu_\tau)\Big\}+h.c.,
 \eea
 with $P_{\chi}=\frac12(I+h_\chi\gamma_5)$ and $h_L=-1,h_R=+1$. Here, $G_F$ denotes the Fermi constant and $V_{cd}$ the corresponding Cabibbo–Kobayashi–Maskawa (CKM) matrix element. The ten Wilson coefficients $C^{V,S}_{cd\tau\chi'\chi}$ and $C^{T}_{cd\tau\chi\chi}$ ( $\chi,\chi'=L,R$), which are in general complex and labeled by the chiralities $\chi,\chi' = L,R$, parametrize the strength of possible physics beyond the Standard Model. Note that tensor operators involving mismatched quark and lepton chiralities vanish identically (see, for example, Refs.~\cite{Mandal:2020htr,Penalva:2021wye}).
 
 It is common to rewrite $H_{\rm eff}$ as
\bea
 H^{\nu_\tau d\to c\tau^-d}_{\rm eff}&=&\frac{2G_F V_{cd}}{\sqrt2}
 \sum_{\chi=L,R}\Big[(\bar c\,[C^V_{cd\tau\chi} \gamma^\mu+h_\chi C^A_{cd\tau\chi}
 \gamma^\mu\gamma_5]\,d)\ (\bar \tau\gamma^\mu P_\chi\nu_\tau)\no
&+&  (\bar c\,[C^S_{cd\tau\chi} +h_\chi C^P_{cd\tau\chi}
\gamma_5]\,d)\ (\bar \tau P_\chi\nu_\tau) +
(\bar c\, C^T_{cd\tau\chi}\sigma^{\mu\nu}[I+h_\chi\gamma_5]\,d)\ (\bar \tau 
\sigma^{\mu\nu}P_\chi\nu_\tau)\Big]
+h.c.,
\label{eq:heffdc}
 \eea
with

\begin{alignat}{2}
C^V_{cd\tau L} &= 1 + C^V_{cd\tau LL} + C^V_{cd\tau RL}   &\qquad
C^A_{cd\tau L} &= 1 + C^V_{cd\tau LL} - C^V_{cd\tau RL}   \nonumber \\
C^S_{cd\tau L} &= C^S_{cd\tau LL} + C^S_{cd\tau RL}       &\qquad
C^P_{cd\tau L} &= C^S_{cd\tau LL} - C^S_{cd\tau RL}       \nonumber \\
C^T_{cd\tau L} &= C^T_{cd\tau LL}                  &\qquad
C^T_{cd\tau R} &= C^T_{cd\tau RR}                 \nonumber \\
C^V_{cd\tau R} &= C^V_{cd\tau LR} + C^V_{cd\tau RR}       &\qquad
C^A_{cd\tau R} &= -(C^V_{cd\tau LR} - C^V_{cd\tau RR})  \nonumber  \\
C^S_{cd\tau R} &= C^S_{cd\tau LR} + C^S_{cd\tau RR}       &\qquad
C^P_{cd\tau R} &= -(C^S_{cd\tau LR} - C^S_{cd\tau RR}) .
\label{eq:wcdctau}
\end{alignat}

For the  $\bar\nu_\tau u\to s\tau^+$ transition the corresponding effective Hamiltonian reads
\bea
 H^{\bar\nu_\tau u\to s\tau^+}_{\rm eff}&=&\frac{2G_F V^*_{us}}{\sqrt2}\sum_{\chi=L,R}
 \Big[(\bar s\,[C^V_{su\tau\chi}\gamma^\mu+h_\chi 
 C^A_{su\tau\chi}
 \gamma^\mu\gamma_5]\,u)\   (\overline{\tau^{\cal C}}\gamma^\mu 
 P_\chi\nu^{\cal C}_\tau)
 +
(\bar s\,[C^S_{su\tau\chi} \no
&+& h_\chi C^P_{su\tau\chi}
\gamma_5]\,(u\   \overline{\tau^{\cal C}} P_\chi\nu^{\cal C}_\tau) +
(\bar s\, C^T_{su\tau\chi}\sigma^{\mu\nu}[I+h_\chi\gamma_5]\,u)\ 
  (\overline{\tau^{\cal C}} 
\sigma^{\mu\nu}P_\chi\nu^{\cal C}_\tau)\Big]
+h.c.,
\label{eq:heffus}
 \eea 
which has been expressed in terms of the charge-conjugate fields
 $\nu^{\cal C}_\tau, \tau^{\cal
 C}$
and where

\begin{alignat}{2}
C^V_{su\tau L}&=-( C^{V*}_{su\tau LR}+C^{V*}_{su\tau RR})   &\qquad
C^A_{su\tau L}&=C^{V*}_{su\tau RR}-C^{V*}_{su\tau LR},   \nonumber \\ 
C^S_{su\tau L}&= C^{S*}_{su\tau LR}+ C^{S*}_{su\tau RR}      &\qquad
C^P_{su\tau L}&=-( C^{S*}_{su\tau LR}- C^{S*}_{su\tau RR})      \nonumber \\
C^T_{su\tau L}&=- C^{T*}_{s\tau RR}                 &\qquad
C^T_{su\tau R}&=- C^{T*}_{su\tau LL}    \nonumber \\ 
C^V_{su\tau R}&=-(1+ C^{V*}_{su\tau LL}+ C^{V*}_{su\tau RL})      &\qquad
C^A_{su\tau R}&=1+ C^{V*}_{su\tau LL}- C^{V*}_{su\tau RL},  \nonumber \\
C^S_{su\tau R}&= C^{S*}_{su\tau LL}+ C^{S*}_{su\tau RL}      &\qquad
 C^P_{su\tau R}&= C^{S*}_{su\tau LL}- C^{S*}_{su\tau RL}
\label{eq:wcsutau}
\end{alignat}

Note the similarity between Eqs.~(\ref{eq:heffdc}) and (\ref{eq:heffus}), and that charge-conjugate fields\footnote{They are defined as $\Psi^{\cal C}=C\bar\Psi^T$, with $C$ the charge-conjugation matrix that satisfies the relation $C\gamma^{\mu T}C^\dagger=-\gamma^\mu$.} play for antiparticles the same role as ordinary fields do for particles. As a result, all expressions obtained for the amplitude of the $\nu_\tau n \to \Lambda_c \tau^-$ reaction can be directly applied to $\bar{\nu}_\tau p \to \Lambda \tau^+$, with the obvious replacements in the values of the CKM matrix elements, masses, form factors, and Wilson coefficients.

%
%
%
%
\section{The  $\nu_\tau n\to
\Lambda_c\tau^-(\pi^-\nu_\tau $) reaction}
\label{sec:Lambdac}
The three dominant $\tau$ decay modes, $\tau \to \pi \nu_\tau$, $\rho \nu_\tau$, and $l\bar\nu_l\nu_\tau$ ($l=e,\mu$), account for more than $70\%$ of the total $\tau$ width. In this work, we focus on the sequential reaction $\nu_\tau n \to \Lambda_c \tau^- (\pi^- \nu_\tau)$, which leads to a charged pion in the final state. 

Accordingly, the cross section for the full process can be expressed as
\be
\sigma=\frac{G_F^2|V_{cd}|^2}{64\pi^5|\vec k\,|}
\int d^3p'\frac{M_{\Lambda_c}}{E'}\int\frac{d^3p_\pi}{E_\pi}
\int\frac{d^3p_{\nu_\tau}}{E_{\nu_\tau}}\delta^{(4)}(k+p-p'-p_\pi-p_{\nu_\tau})
\sum_{\chi=L,R}\rho_{\chi\chi}\sum_{a,b}
W^{ab}_\chi{\cal L}_{\chi ab}, \label{eq:WLsigma}
\ee
where $k,\,p ,\,p',\,p_\pi$, and $p_{\nu_\tau}$ denote the four-momenta of the initial $\nu_\tau$, the nucleon, the final hyperon, the pion, and the final $\nu_\tau$, respectively, with $\rho_{LL}$ and $\rho_{RR}$ representing the diagonal components of the incident neutrino density matrix.\footnote{The off-diagonal contributions vanish when neutrino mass terms are neglected.}  At energies much larger than its mass, left- and right-handed chirality correspond to negative and positive helicity, respectively, an identification used throughout this analysis. Even with potential NP, neutrino production is expected to remain SM-dominated, so the typical case is a fully left-handed neutrino beam ($\rho_{LL}=1$, $\rho_{RR}=0$), while for an antineutrino beam the opposite applies ($\rho_{LL}=0$, $\rho_{RR}=1$). In Eq.~\eqref{eq:WLsigma}, the indices $a$ and $b$ denote the three allowed Dirac structures: (pseudo-)vector, (pseudo-)scalar, and tensor. The associated dimensionless hadron tensors $W^{ab}_\chi$ read
\bea
&&W^{ab}_\chi(p,q)=\frac12\sum_{r,r'}\langle\Lambda_c;p',r'|\bar c( 0)O^a_{H\chi} d(0)|n;p,r\rangle
\langle\Lambda_c;p',r'|\bar c( 0)O^b_{H\chi} d(0)|n;p,r\rangle^*,\label{eq:ht}\\
&&O^a_{H\chi}=C^V_{dc\tau\chi}\gamma^\mu+h_\chi C^A_{dc\tau\chi}\gamma^\mu\gamma_5,\quad
C^S_{dc\tau\chi}+h_\chi C^P_{dc\tau\chi}\gamma_5, \quad 
C^T_{dc\tau\chi}\sigma^{\mu\nu}(I+h_\chi\gamma_5),
\eea 
where we average (sum) over the spins of the initial (final) baryon. The resulting expressions are fully general and can be read off from Appendix C of Ref.~\cite{Penalva:2021wye}, where they are given for a similar 
 $H_b \to H_c$ transition. Throughout this work we adopt the convention 
$q=p-p'=k'-k$, corresponding to minus the lepton four-momentum transfer. This choice allows to use directly the results derived for decay processes in Appendix C of Ref.~\cite{Penalva:2021wye} for the hadronic part of the amplitude.~\footnote{In Ref.~\cite{Penalva:2021wye}, $M$ denotes the mass of the initial hadron, which in the present case corresponds to $M_n$.}  These tensors are constructed from hadron masses and momenta and are expressed in terms of dimensionless structure functions $\widetilde W_\chi$, which depend on the form factors,\footnote{The form-factor parametrization adopted for a $1/2^+ \to 1/2^+$ transition is given in Ref.~\cite{Hernandez:2025snr}.} the Wilson coefficients, and the masses involved. Explicit expressions for $\widetilde W_\chi$ are available for several transitions, including $1/2^+ \to 1/2^+$~\cite{Penalva:2020xup}, which is relevant for this work, as well as $0^- \to 0^-,1^-$~\cite{Penalva:2020ftd} and $1/2^+ \to 1/2^-,3/2^-$~\cite{Du:2022ipt}. Although their explicit form depends on the specific transition, the formalism remains fully general once the structure functions are specified. 

The tensor for the two neutrinos and the final pion is given by the product of lepton currents $l_{\chi a}$, constructed with the Dirac matrices $\Gamma_a = \gamma_\mu, I $ and $ \sigma_{\mu\nu}$, reads (see Appendix~\ref{app:lepton} for details)
\begin{equation}
{\cal L}_{\chi ab}= l_{\chi a} l^{\dagger}_{\chi b}={\cal B}(\tau\to\pi\nu_\tau)
\frac{32\pi^2 m_\tau^2 }{(m^2_\tau-m^2_\pi)^2}{\rm Tr}\,[\slashed{p}_{\nu_\tau} P_R(\slashed{k}'+m_\tau)
\Gamma_a \frac1{2}P_\chi\slashed{k}\gamma^0\Gamma^\dagger_b\gamma^0
(\slashed{k}'+m_\tau)P_L]\,\delta(k^{\prime2}-m^2_\tau),\label{eq:calL}
\end{equation}
with  $k' = k + p - p' = p_\pi + p_{\nu_\tau}$ the intermediate $\tau$ four-momentum and ${\cal B}(\tau\to\pi\nu_\tau)$ the branching ratio for the  $\tau\to\pi\nu_\tau$ decay.

Equation~\eqref{eq:WLsigma} can then be written as (see also Appendix~\ref{app:lepton})
\bea
\sigma
&=&{\cal B}(\tau\to\pi\nu_\tau)\frac{ m^2_\tau M_{\Lambda_c}G_F^2|V_{cd}|^2}
{4\pi^3|\vec k\,|(m^2_\tau-m^2_\pi)}
\int\frac{d^3p'}{E'}\int\frac{d^3p_\pi}{ E_\pi}\int\frac{d^3p_{\nu_\tau}}
{ E_{\nu_\tau}}
\delta^{(4)}(k+p-p'-p_\pi-p_{\nu_\tau})\nonumber\\
&\times&\delta((k+p-p')^2-m^2_\tau)\Big[\sum_{\chi=L,R}\rho_{\chi\chi}\sum_{a,b}W^{ab}_\chi L_{\chi ab}\Big]
\Big(1+\frac{2m_\tau}{m^2_\tau-m^2_\pi}\,p_\pi\cdot{\cal P}
\Big),\label{eq:sec1}
\eea
where we have introduced the  $\tau$ polarization four-vector ${\cal P}^\mu$ and the purely leptonic tensors $L_{\chi ab}(k,k')$,
\bea
L_{\chi ab}(k,k')=\frac14{\rm Tr}[(\slashed{k'}+m_\tau)\Gamma_a(I+h_\chi\gamma_5)\slashed{k}
\gamma^0\Gamma^\dagger_b\gamma^0],
\eea
These latter tensors enter the evaluation of the $\nu_\tau n \to \Lambda_c \tau^-$ transition and their expressions can be obtained from Appendix B of Ref.~\cite{Penalva:2021wye} by taking twice the unpolarized part of the lepton tensors. On the other hand, the Lorentz components of  ${\cal P^\mu}$ are
\bea
{\cal P}^\mu=\frac{\sum_{\chi=L,R}\rho_{\chi\chi}\sum_{a,b}W^{ab}_\chi\ {\rm Tr\,}
[(\slashed{k}'+m_\tau)
\Gamma_a
\frac12P_\chi\slashed{k}\gamma^0\Gamma^\dagger_b\gamma^0(\slashed{k}'+m_\tau)
\gamma_5\gamma^\mu]}{2m_\tau\sum_{\chi'=L,R}\rho_{\chi'\chi'}\sum_{c,d}W^{cd}_{\chi'} L_{\chi' cd}}. \label{eq:P}
\eea
The $\tau$ polarization vector can be written as a linear combination of the four-momenta
$p^\mu_\perp$ and $q^\mu_\perp$, defined as
$\ell_\perp^\mu = \ell^\mu - (\ell \cdot k'/m_\tau^2)\,k'^\mu$ with $\ell = p,q$,
together with the pseudo-vector
$\epsilon^{\mu k' q p} \equiv \epsilon^{\mu\nu\rho\sigma} k'_\nu q_\rho p_\sigma$
(where $\epsilon_{0123}=+1$).
The corresponding coefficients depend on the particle masses, the scalar product
$(p \cdot k)$, and the variable
$\omega = (p \cdot p')/(M_n M_{\Lambda_c})$, related to the squared momentum transfer by $
q^2 = M_n^2 + M_{\Lambda_c}^2 - 2 M_n M_{\Lambda_c}\,\omega$ \cite{Penalva:2021gef}. Further details are provided in Appendix~\ref{app:P}.

\begin{figure}[htb]
\begin{center}
\includegraphics[height=.35\textwidth]{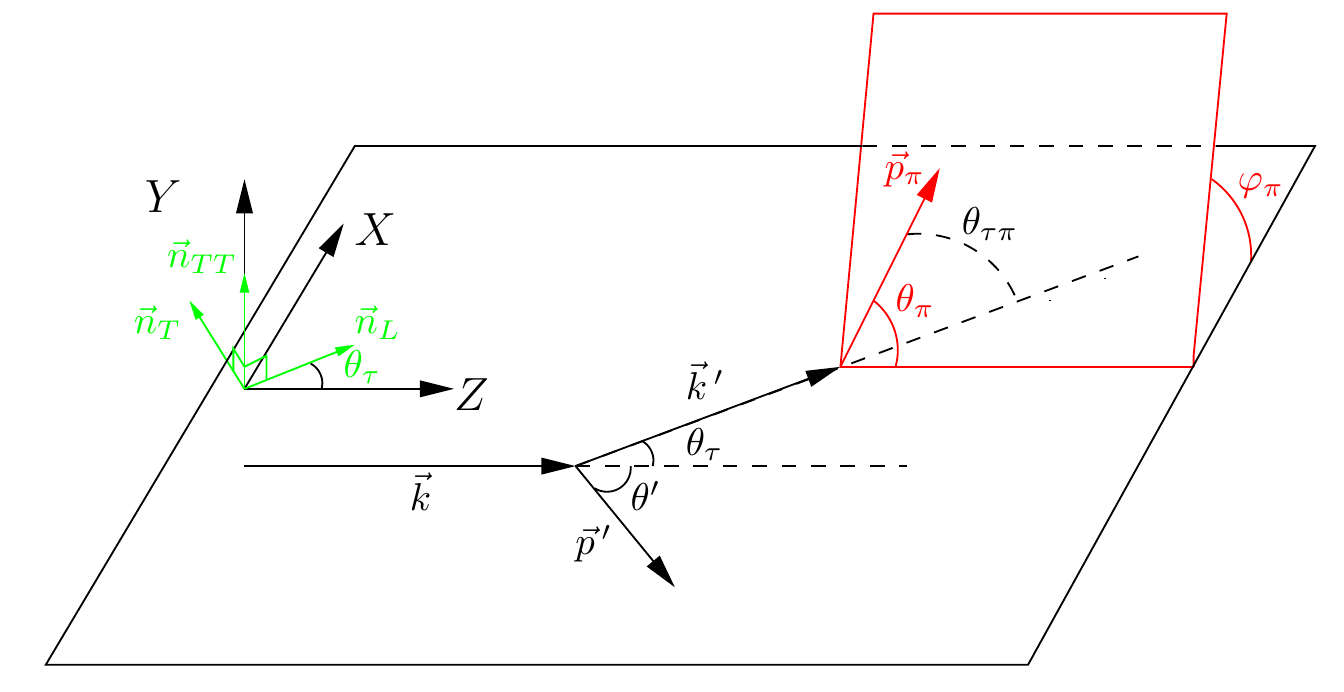}
\caption{Neutron-LAB kinematics for $\nu_\tau n \to \Lambda_c \tau^- 
(\pi^- \nu_\tau)$, showing the three-momenta $\vec{k}$, $\vec{k}'$, $\vec{p},'$ 
and $\vec{p}_\pi$ of the incoming neutrino, intermediate $\tau$, outgoing 
$\Lambda_c$, and pion. We also display the longitudinal unit vector 
$\vec n_L = \vec k'/|\vec k'|$ and the transverse unit vectors $\vec n_T$ 
and $\vec n_{TT}$ entering the definition of $N_L^\mu$, $N_T^\mu$, and 
$N_{TT}^\mu$ in Eq.~\eqref{eq:nvecpol}.}  
\label{fig:kin}
\end{center}
\end{figure}
It is customary to decompose the polarization four-vector ${\cal P}^\mu$ as
\bea
{\cal P}^\mu={\cal P}_L N_L^\mu+{\cal P}_T N_T^\mu+{\cal P}_{TT} N_{TT}^\mu \label{eq:plttt}
\eea
where\footnote{These three spatial four-vectors, together with $k^{\prime\mu}/m_\tau$, form a basis of four-vector space. Note, however, that ${\cal P}^\mu \perp k^{\prime\mu}$.}
\bea
N^\mu_L=\Big(\frac{|\vec k\,'|}{m_\tau},\frac{k^{\prime0}\,\vec k\,'}
{m_\tau|\vec k\,'|}\Big),\ N^\mu_T=\Big(0,\frac{(\vec k\times\vec k\,')
\times\vec k\,'}
{|(\vec k\times\vec k\,')\times\vec k\,'|}\Big),\ 
N^\mu_{TT}=\Big(0,\frac{\vec k\times\vec k\,'}
{|\vec k\times\vec k\,'|}\Big) \label{eq:nvecpol}
\eea
The above decomposition defines the longitudinal (${\cal P}_L$) and transverse (${\cal P}_T$, ${\cal P}_{TT}$) components of the polarization vector; their explicit expressions are given in Appendix~\ref{app:P} (see also Fig.~\ref{fig:kin}). The ${\cal P}_{TT}$ component\footnote{Projection of ${\cal P}^\mu$ along the direction normal to the plane spanned by the incoming neutrino and the outgoing $\tau$ (or hyperon $\Lambda_c$).}  arises solely from the $\epsilon^{\mu k' q p}$ term of ${\cal P}^\mu$, which is not invariant under time reversal, and its nonzero value requires some Wilson coefficients to be complex~\cite{Penalva:2021gef}. Consequently, an observation of ${\cal P}_{TT} \neq 0$ would provide a clear signal of CP violation and BSM physics. Contributions to ${\cal P}_{TT}$ stem from the interference of vector–axial with scalar–pseudoscalar terms, scalar–pseudoscalar with tensor terms, and vector–axial with tensor terms. Since vector–axial contributions are already present in the SM, at least one of the Wilson coefficients $C^S_\chi$, $C^P_\chi$, or $C^T_\chi$ must be nonzero.

Exploiting rotational invariance and the delta functions to perform part of the phase-space integrations in Eq.~\eqref{eq:sec1}, we obtain (see Appendix~\ref{app:kin} and Fig.~\ref{fig:kin}),
\bea
\sigma
&=&{\cal B}(\tau\to\pi\nu_\tau)\frac{ m^2_\tau M_{\Lambda_c}G_F^2|V_{cd}|^2}
{2\pi^2|\vec k\,|^2(m^2_\tau-m^2_\pi)}
\int_{E'_-}^{E'_+} \frac{dE'}{\sqrt{(|\vec k\,|+M_n-E')^2-m^2_\tau}}\int_{E^-_{\pi}(E')}^{E^+_{\pi}(E')}dE_\pi
\int_{\cos(\theta^0_{\tau\pi}+\theta^0_\tau)}^{\cos(\theta^0_{\tau\pi}-\theta^0_\tau)}
d\cos\theta_\pi \no
&\times&\frac{\sum_{\chi=L,R}\rho_{\chi\chi}\sum_{a,b}W^{ab}_\chi L_{\chi ab}}{\sqrt{(\cos\theta_\pi-\cos(\theta^0_{\tau\pi}+\theta^0_\tau))
(\cos(\theta^0_{\tau\pi}-\theta^0_\tau)-\cos\theta_\pi)}}\Bigg\{1+
\frac{2m_\tau}{m^2_\tau-m^2_\pi}\,\Big(\frac{{\cal P}_L}{m_\tau}[
E_\pi\sqrt{(|\vec k\,|+M_n-E')^2-m^2_\tau}\no
&-&|\vec p_\pi|(|\vec k\,|+M_n-E')\cos\theta^0_{\tau\pi}]+\frac{{\cal P}_T|\vec p_\pi|}{\sin\theta^0_\tau}[
\cos\theta_\pi-\cos\theta^0_\tau\cos\theta^0_{\tau\pi}]\Big)
\Bigg\},\label{eq:sig1}
\eea
where $\theta'_0$, $\theta^0_{\tau}$, $\varphi^0_\pi$, and $\theta^0_{\tau\pi}$ are the angles fixed by momentum-conservation  and depend on the remaining integration variables (Eq.~\eqref{eq:aux1}), while the integration limits for $E_{\Lambda_c}$ and $E_\pi$ are given in Eqs.~\eqref{eq:Eprimelimits} and \eqref{eq:Epionlimits}. 

We see in Eq.~\eqref{eq:sig1} that all the information on ${\cal P}_{TT}$ is  lost. This results from the integration over the pion azimuthal angle: ${\cal P}_{TT}$ always appears multiplied by $\sin\varphi_\pi$ (see Eq.~\ref{eq:pttANDsinphipi}), and the delta-function constraints yield two solutions for $\varphi_\pi$ with the same cosine but opposite sine. That is, if $\varphi^0_\pi \in [0,\pi]$ satisfies momentum conservation, then $(2\pi - \varphi^0_\pi) \in [\pi,2\pi]$ does as well. However, by separately counting events with pions emitted below or above the plane defined by the incoming neutrino and the intermediate $\tau$ (or equivalently in this case, the outgoing hyperon $\Lambda_c$, see Fig.~\ref{fig:kin}), one can construct an asymmetry
\bea
\Delta\sigma&=&\sigma(0<\varphi_\pi<\pi)-
\sigma(\pi<\varphi_\pi<2\pi)\no
&=&{\cal B}(\tau\to\pi\nu_\tau)\frac{ m^2_\tau M_{\Lambda_c}G_F^2|V_{cd}|^2}
{2\pi^2|\vec k\,|^2(m^2_\tau-m^2_\pi)}
\int_{E'_-}^{E'_+} \frac{dE'}{\sqrt{(|\vec k\,|+M_n-E')^2-m^2_\tau}}\int_{E^-_{\pi}(E')}^{E^+_{\pi}(E')}dE_\pi
\int_{\cos(\theta^0_{\tau\pi}+\theta^0_\tau)}^{\cos(\theta^0_{\tau\pi}-\theta^0_\tau)}
d\cos\theta_\pi \no
&\times&\frac{\sum_{\chi=L,R}\rho_{\chi\chi}\sum_{a,b}W^{ab}_\chi L_{\chi ab}}{\sqrt{(\cos\theta_\pi-\cos(\theta^0_{\tau\pi}+\theta^0_\tau))
(\cos(\theta^0_{\tau\pi}-\theta^0_\tau)-\cos\theta_\pi)}}
\,\frac{2m_\tau}{m^2_\tau-m^2_\pi}\Big(-{\cal P}_{TT}\, |\vec p_\pi|
\,|\sin\varphi^0_\pi|\,\sin\theta_\pi
\Big)\label{eq:sig2}
\eea
which is sensitive to ${\cal P}_{TT}$. As discussed above, a nonzero value of this asymmetry would signal the presence of CP-violating NP beyond the SM.

Note that for the cross section $\sigma$, since the polarization components ${\cal P}_{L,T}$ are independent of the pion kinematics, the integration over $\cos\theta_\pi$ can be performed using Eq.~(F.6) of Ref.~\cite{Penalva:2020xup}.
\bea
\sigma
&=&{\cal B}(\tau\to\pi\nu_\tau)\frac{ M_{\Lambda_c}G_F^2|V_{cd}|^2}
{2\pi|\vec k\,|^2}
\int_{E'_-}^{E'_+} \frac{dE'}{ E^+_{\pi}(E')-E^-_{\pi}(E')}\Big[\sum_{\chi=L,R}\rho_{\chi\chi}\sum_{a,b}W^{ab}_\chi L_{\chi ab}\Big]
\no
&\times&\int_{E^-_{\pi}(E')}^{E^+_{\pi}(E')}dE_\pi \Bigg\{1+
\frac{E^+_{\pi}(E')+E^-_{\pi}(E')
-2  E_\pi
}{ E^+_{\pi}(E')-E^-_{\pi}(E')}\,{\cal P}_L(E') \label{eq:integrada}
\Bigg\}.
\eea
We see that sensitivity to ${\cal P}_T$ is further lost. Moreover, the term proportional to ${\cal P}_L$ vanishes upon integration over $E_\pi$, leading to
\bea
\sigma={\cal B}(\tau\to\pi\nu_\tau)\Bigg(\frac{ M_{\Lambda_c}G_F^2|V_{cd}|^2}
{2\pi|\vec k\,|^2}
\int_{E'_-}^{E'_+} {dE'}\sum_{\chi=L,R}\rho_{\chi\chi}\sum_{a,b}W^{ab}_\chi L_{\chi ab}\Bigg),
\eea
which simply reproduces  $\sigma[\nu_\tau n\to \Lambda_c\tau^-(\pi^-\nu_\tau)]= {\cal B}(\tau\to\pi\nu_\tau)\ \sigma(\nu_\tau n\to
\Lambda_c\tau^-)$.

Following the strategy of Ref.~\cite{Penalva:2022vxy}, the statistical sensitivity of the pion energy and angular distributions is improved by first integrating over the hyperon energy $E'$. This procedure leads to the double-differential distributions 
$d^2\sigma/(dE_\pi d\cos\theta_\pi)$ and $d^2\Delta\sigma/(dE_\pi d\cos\theta_\pi)$, whose explicit expressions are given in Eqs.~\eqref{eq:sec3} and \eqref{eq:asymmetry} of Appendix~\ref{app:kin}. Further integration over either charged-pion variable yields the corresponding single-differential cross sections.

The cross section for the analogous process $\nu_\tau n \to \Lambda_c \tau^- (\rho^- \nu_\tau)$ can be obtained through the straightforward replacements $m_\pi \to m_\rho$ and ${\cal B}(\tau \to \pi \nu_\tau) \to {\cal B}(\tau \to \rho \nu_\tau)$, together with a rescaling of the polarization term in Eq.~\eqref{eq:sig1} by the $\rho$ analyzing power, $(m_\tau^2 - 2 m_\rho^2)/(m_\tau^2 + 2 m_\rho^2) < 1$~\cite{Penalva:2021wye}. Reference~\cite{Penalva:2021wye} also provides the necessary ingredients to address the four-particle final-state processes $\nu_\tau n \to \Lambda_c \tau^- (\mu^- \bar\nu_\mu \nu_\tau)$ and $\nu_\tau n \to \Lambda_c \tau^- (e^- \bar\nu_e\nu_\tau)$. Analogous considerations apply to the antineutrino-induced reaction $\bar\nu_\tau p \to \Lambda \tau^+ (\pi^+ \bar\nu_\tau)$.

\section{The $\bar\nu_\tau p\to\Lambda \tau^+(\pi^+\bar\nu_\tau)$ reaction}
\label{sec:Lambda}
 In addition to the obvious changes in the values of the CKM matrix elements, masses, form factors, Wilson coefficients, and density-matrix elements entering the evaluation of $\sum_{\chi=L,R}\rho_{\chi\chi}\sum_{a,b}W^{ab}_\chi 
L_{\chi ab}$—as discussed in Appendix~\ref{app:anti} when deriving the $\bar\nu_\tau p\to\Lambda \tau^+(\pi^+\bar\nu_\tau)$ distributions from the results of the previous section—one must also implement the additional modification
\bea
\Big(1+\frac{2m_\tau}{m^2_\tau-m^2_\pi}\,p_\pi\cdot{\cal P}\Big)\longrightarrow
\Big(1-\frac{2m_\tau}{m^2_\tau-m^2_\pi}\,p_\pi\cdot{\cal P}\Big)
\label{eq:change}
\eea
with ${\cal P}$ defined exactly as before but evaluated using the $\Lambda$  and proton masses, the corresponding Wilson coefficients from Eq.~(\ref{eq:heffus}), the appropriate form factors for the $p\to \Lambda$ transition, and the density-matrix elements corresponding to the antineutrino beam.

For this transition, we will use two different sets of form factors. First, we consider the
$\Lambda\to p$ semileptonic decay helicity form factors computed in Ref.~\cite{Bacchio:2025auj} using lattice QCD (LQCD). These results must be analytically continued to the region of large negative $q^2$, which could introduce uncertainties in the theoretical evaluation of the various observables considered here. The second set corresponds to the phenomenological determination provided in Ref.~\cite{Mintz:2004hm}. In both cases, only the vector and axial form factors are given; however, the scalar and pseudoscalar form factors can be obtained using the equations of motion\footnote{The corresponding expressions for  the $F_S$ and $F_P$ form factors in the $n\to\Lambda_c$ transition are given in Appendix A of Ref.~\cite{Hernandez:2025snr} and can be applied here with the appropriate substitutions. }. 
\section{Results}
In what follows, we shall assume the neutrino/antineutrino beam to be fully left/right-handed.
\label{sec:results}
\subsection{$\nu_\tau n\to\Lambda_c \tau^-(\pi^-\nu_\tau)$}
In this subsection, we present results for the 
$\nu_\tau n \to \Lambda_c \tau^- (\pi^- \nu_\tau)$ reaction, including the 
pion double-differential cross section $d^2\sigma/(dE_\pi\, d\cos\theta_\pi)$, 
the single-differential cross sections $d\sigma/dE_\pi$ and 
$d\sigma/d\cos\theta_\pi$, as well as the angular CP-asymmetry, $\Delta\sigma$, defined in Eq.~\eqref{eq:sig2}.
In all cases, we use the LQCD form factors computed in Ref.~\cite{Meinel:2017ggx}. As discussed in Ref.~\cite{Hernandez:2025snr}, these form factors must be analytically continued from the $q^2 > 0$ region of the $\Lambda_c\to n$ decay to the $q^2 < 0$ relevant for the $\nu_\tau n \to \Lambda_c\tau$
reaction. This extrapolation outside the fitted region introduces significant uncertainties. The impact of these uncertainties on the 
$\nu_\tau n \to \Lambda_c\tau$ cross section was analyzed in Ref.~\cite{Hernandez:2025snr}, with the conclusion that they limit the precision of theoretical predictions. The results presented below have been evaluated using the central values of the form factors from Ref.~\cite{Meinel:2017ggx} and, in the absence of a more reliable determination in the large negative $q^2$ region, should be considered indicative of the expected behavior of these differential cross sections.

\begin{figure}[htb]
\begin{center}
\includegraphics[height=.45\textwidth]{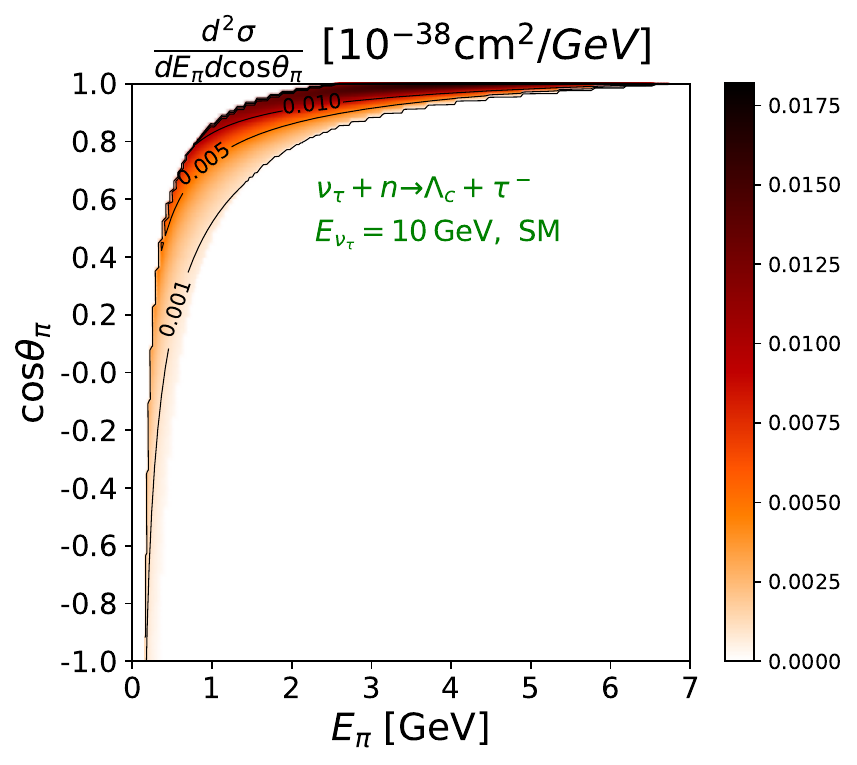}
\caption{ Double-differential cross section $d^2\sigma/(dE_\pi\, d\cos\theta_\pi)$ for the
$\nu_\tau n \to \Lambda_c \tau^- (\pi^- \nu_\tau)$ reaction at $E_{\nu_\tau}=10\,\mathrm{GeV}$,
evaluated within the SM for a fully left-handed polarized $\nu_\tau$ beam. The form factors have been taken from Ref.~\cite{Meinel:2017ggx}.}  
\label{fig:d2sigma}
\end{center}
\end{figure}

In Fig.~\ref{fig:d2sigma}, we present the double-differential cross section
$d^2\sigma/(dE_\pi\, d\cos\theta_\pi)$ evaluated at $E_{\nu_\tau}=10\,\mathrm{GeV}$
within the SM. The available pion phase space is strongly restricted, forming a
narrow kinematical band in the $(E_\pi,\cos\theta_\pi)$ plane. The cross section
exhibits a pronounced accumulation of strength in the forward region, with pions
predominantly emitted at small angles relative to the incoming neutrino
direction, i.e., $\cos\theta_\pi \approx 1$.

\begin{figure}[htb]
\begin{center}
\rotatebox{270}{\includegraphics[height=.45\textwidth]{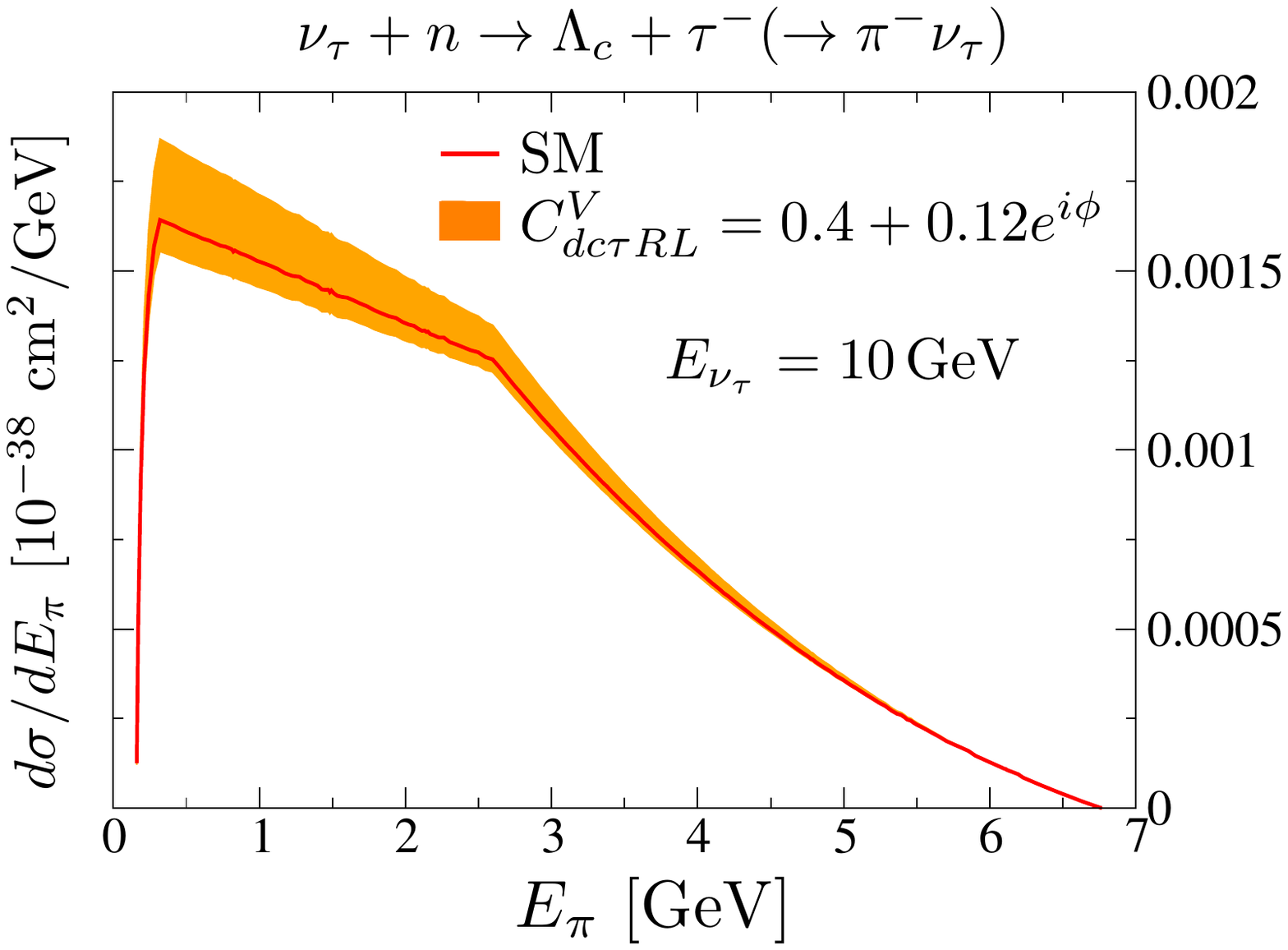}}
\rotatebox{270}{\includegraphics[height=.45\textwidth]{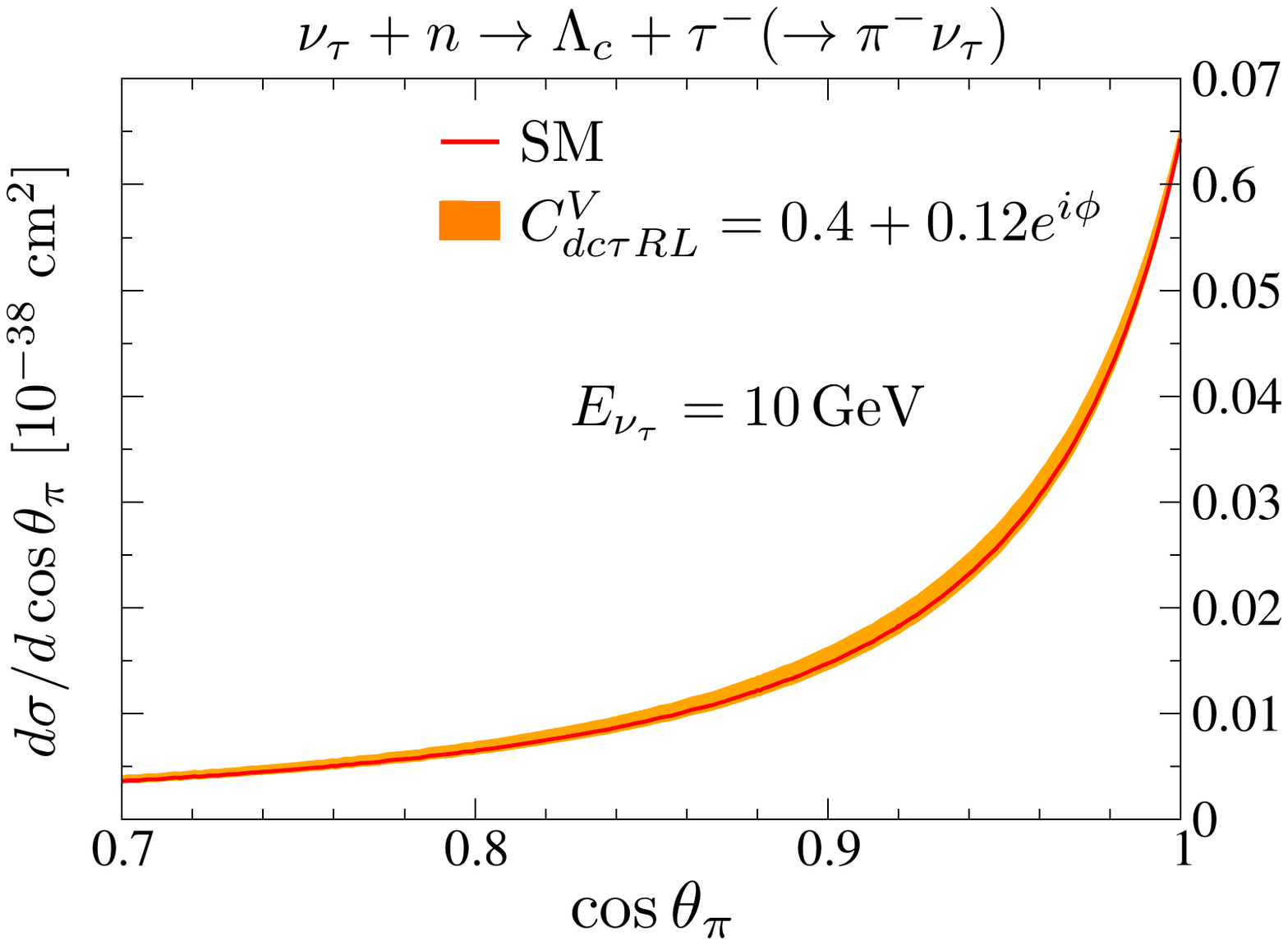}}
\caption{Differential cross sections $d\sigma/dE_\pi$ (left) and $d\sigma/d\cos\theta_\pi$ (right) for the
$\nu_\tau n \to \Lambda_c \tau^- (\pi^- \nu_\tau)$ reaction evaluated at
$E_\nu = 10\,\mathrm{GeV}$, for a fully left-handed polarized $\nu_\tau$ beam.
The SM prediction (red curve) is shown together with results including NP effects (orange-shaded band) induced by a nonzero Wilson coefficient $C^V_{dc\tau RL}=0.04+0.12e^{i\phi}$. In both panels, the upper limit of the band is reached for
$\phi=0$, while for the lower limit $\phi=\pi$.}  
\label{fig:dsigdepi}
\end{center}
\end{figure}

This feature is more clearly exposed in the right panel of
Fig.~\ref{fig:dsigdepi}, which shows the single-differential distribution
$d\sigma/d\cos\theta_\pi$ and reveals a strong forward peaking. Complementarily,
the left panel of Fig.~\ref{fig:dsigdepi} displays the $d\sigma/dE_\pi$
distribution, where the cross section is seen to be concentrated at low pion
energies—an effect that is not immediately evident from the two-dimensional
representation in Fig.~\ref{fig:d2sigma}.

In Fig.~\ref{fig:dsigdepi}, we also illustrate the impact of allowing for a nonzero
$C^V_{dc\tau RL}$ Wilson coefficient. As discussed in
Ref.~\cite{Hernandez:2025snr}, this is the only Wilson coefficient that is not
strongly constrained to small values by the combined analysis of Drell--Yan data
and purely leptonic $D$-meson decays performed in
Ref.~\cite{Fuentes-Martin:2020lea}. Consequently, it is the only coefficient
capable of inducing sizable deviations from  SM predictions.
Following the same conservative strategy as in Ref.~\cite{Hernandez:2025snr}, we assume that deviations from the SM remain moderate and take  $|C^V_{dc\tau RL}-0.04| \simeq 0.12$, ensuring that $|C^V_{dc\tau RL}| = \mathcal{O}(10^{-1})$ while staying consistent with the constraints derived in Ref.~\cite{Fuentes-Martin:2020lea}. 
The resulting impact is illustrated in Fig.~\ref{fig:dsigdepi} as a band. 
We see that this Wilson coefficient can either enhance or suppress the differential cross sections, while largely preserving the overall shapes of the distributions.

We now turn to the CP asymmetry, $\Delta\sigma$, defined in Eq.~\eqref{eq:sig2}. As mentioned above, the asymmetry arises from the interference  between vector-axial terms with scalar-pseudoscalar or tensor terms, as well as between scalar-pseudoscalar 
and tensor terms. For the asymmetry to be nonzero, at least one of the Wilson coefficients 
$C^S_{dc\tau L}$, $C^P_{dc\tau L}$, or $C^T_{dc\tau L}$ must be nonzero, and at least one of the involved coefficients must be complex.

Since $C^V_{dc\tau L}$ and $C^A_{dc\tau L}$ are already $\mathcal{O}(1)$ within the SM and BSM corrections are expected to be small, 
the largest contributions to the asymmetry come from a nonzero imaginary part in 
$C^S_{dc\tau L}$, $C^P_{dc\tau L}$, or $C^T_{dc\tau L}$, i.e., from the interference of vector-axial with scalar-pseudoscalar or tensor terms.
In the following, we assume that only one of $C^S_{dc\tau LL}$, $C^S_{dc\tau RL}$, or $C^T_{dc\tau LL}$ is nonzero at a time and purely imaginary. 
This approach allows us to illustrate the possible magnitude of the CP-asymmetry if one of these Wilson coefficients is complex. 
The results for $E_{\nu_\tau} = 10\,\mathrm{GeV}$ are presented in Table~\ref{tab:asy}. 
The modulus of each coefficient is constrained using the 95\% confidence level limits from Ref.~\cite{Fuentes-Martin:2020lea}, derived from Drell-Yan data.

As expected, the resulting asymmetry values are very small compared to the cross section. 
Nevertheless, any nonzero asymmetry, however tiny, would indicate the presence of CP-violating  NP. 
The table also reports the fractional change in the total cross section, which is similarly found to be small.
\begin{table}[h!]
\begin{center}
\begin{tabular}{|c|c|c|}\hline
      $C^S_{dc\tau LL}$      &  $\frac{\Delta\sigma}{\sigma}(\%)$ & 
      $\frac{\sigma-\sigma_{\rm SM}}{\sigma_{\rm SM}}(\%)$\\\hline
$0.01i$&$-0.13$& $2.8\times10^{-3}$\\ 
$0.03i$&$-0.39$&$2.6\times10^{-2}$\\\hline 
\end{tabular}
\begin{tabular}{|c|c|c|}\hline
      $C^S_{dc\tau RL}$      &  $\frac{\Delta\sigma}{\sigma}(\%)$ & 
      $\frac{\sigma-\sigma_{\rm SM}}{\sigma_{\rm SM}}(\%)$\\\hline
$0.01i$&$-0.072$& $2.8\times10^{-3}$\\ 
$0.03i$&$-0.22$&$2.6\times10^{-2}$\\\hline \end{tabular}
\begin{tabular}{|c|c|c|}\hline
      $C^T_{dc\tau LL}$      &  $\frac{\Delta\sigma}{\sigma}(\%)$ & 
      $\frac{\sigma-\sigma_{\rm SM}}{\sigma_{\rm SM}}(\%)$\\\hline
$0.001i$&$-4.4\times10^{-3}$& $6.7\times10^{-4}$\\ 
$0.007i$&$-3.1\times10^{-2}$&$6.5\times10^{-2}$\\\hline \end{tabular}
\end{center}
\caption{CP asymmetry $\Delta\sigma$ [Eq.~\eqref{eq:sig2}] and total cross-section variations $(\sigma-\sigma_{\rm SM})$ for the 
$\nu_\tau n \to \Lambda_c \tau^- (\pi^- \nu_\tau)$ reaction, evaluated at 
$E_\nu = 10\,\mathrm{GeV}$ for a fully left-handed $\nu_\tau$ beam. The SM total cross section is $\sigma_{\rm SM}=\left (0.0056^{+0.0049}_{-0.0018}\right)\times10^{-38}$\,cm$^2$, where the uncertainties take into account both the statistical and systematics errors in the determination of the form factors. The central value is used in this work. The NP effects are generated by assuming that one of the Wilson coefficients $C^S_{dc\tau LL}$, $C^S_{dc\tau RL}$, or $C^T_{dc\tau LL}$ is nonzero and purely imaginary. 
See text for further details.}
\label{tab:asy}
\end{table} 
These predictions provide an estimate of what to expect if the reaction occurs on a deuteron. 
For heavier nuclei, a proper treatment of nuclear medium effects is required, which will be addressed in a future work.

\subsection{$\bar\nu_\tau p\to\Lambda \tau^+(\pi^+\bar\nu_\tau)$}

Here, we present nucleon-level results for the 
$\bar\nu_\tau p \to \Lambda \tau^+ (\pi^+ \bar\nu_\tau)$ reaction. As in the $n \to \Lambda_c$ case, we show the pion double-differential cross section 
$d^2\sigma/(dE_\pi\, d\cos\theta_\pi)$, the single-differential cross sections 
$d\sigma/dE_\pi$ and $d\sigma/d\cos\theta_\pi$, as well as the angular  CP-asymmetry $\Delta\sigma$, all evaluated at 
$E_{\bar\nu_\tau} = 5\,\mathrm{GeV}$.

\begin{figure}[htb]
\begin{center}
\includegraphics[height=.4\textwidth]{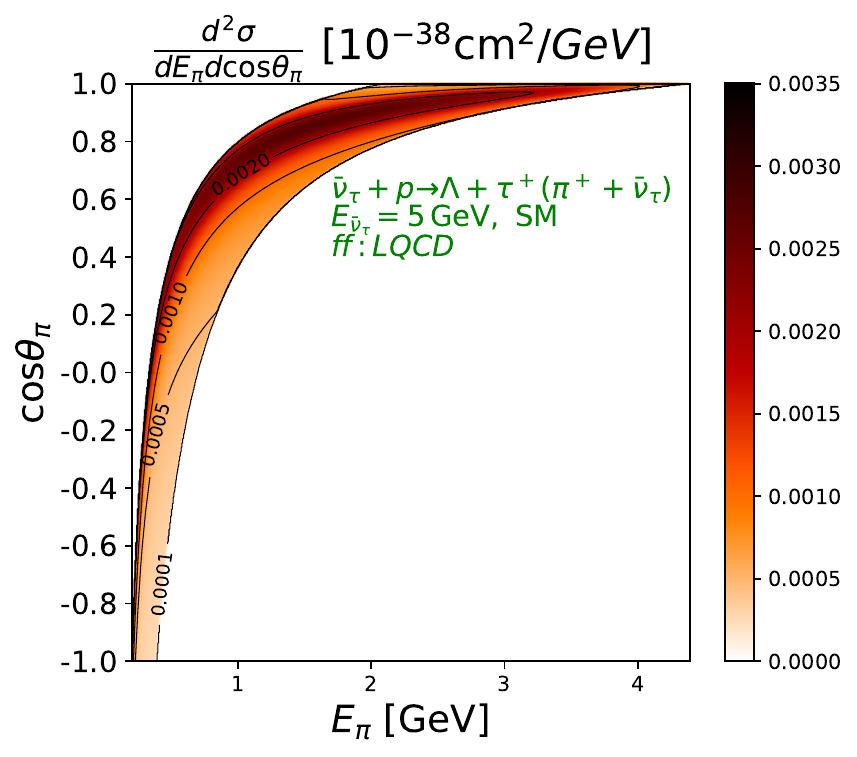}\ 
\includegraphics[height=.4\textwidth]{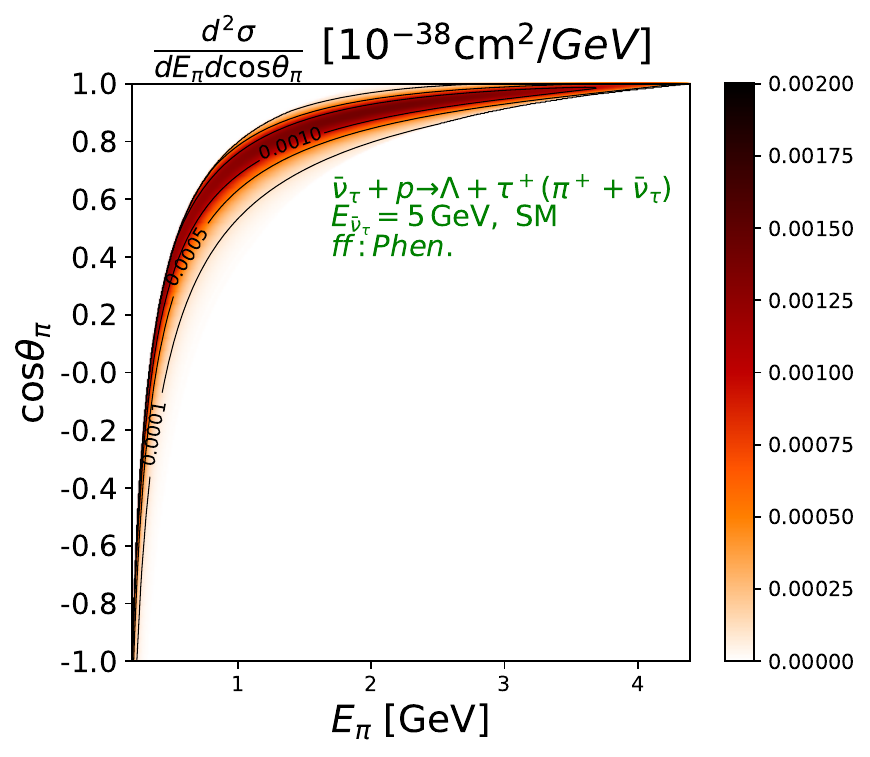}\caption{ Double-differential cross section $d^2\sigma/(dE_\pi\, d\cos\theta_\pi)$ 
for the $\bar\nu_\tau p \to \Lambda \tau^+ (\pi^+ \bar\nu_\tau)$ reaction, 
evaluated at $E_{\bar\nu_\tau} = 5\,\mathrm{GeV}$ within the SM for a fully 
right-handed $\bar\nu_\tau$ beam. The left panel shows results using the 
LQCD form factors from Ref.~\cite{Bacchio:2025auj}, while the right panel is 
evaluated with the phenomenological form factors from Ref.~\cite{Mintz:2004hm}.}  
\label{fig:d2sigmalambda}
\end{center}
\end{figure}

In Fig.~\ref{fig:d2sigmalambda}, we display the SM double-differential cross section
$d^2\sigma/(dE_\pi\, d\cos\theta_\pi)$. The left and right panels are obtained using,
respectively, the LQCD form factors from Ref.~\cite{Bacchio:2025auj} and the
phenomenological form factors of Ref.~\cite{Mintz:2004hm}. As before, the
allowed pion phase space is highly constrained, resulting in a narrow
kinematical band. Clear differences between the two form-factor prescriptions
are already apparent, impacting both the shape of the distributions and the
overall magnitude of the cross section.

\begin{figure}[htb]
\begin{center}
\rotatebox{270}{\includegraphics[height=.45\textwidth]{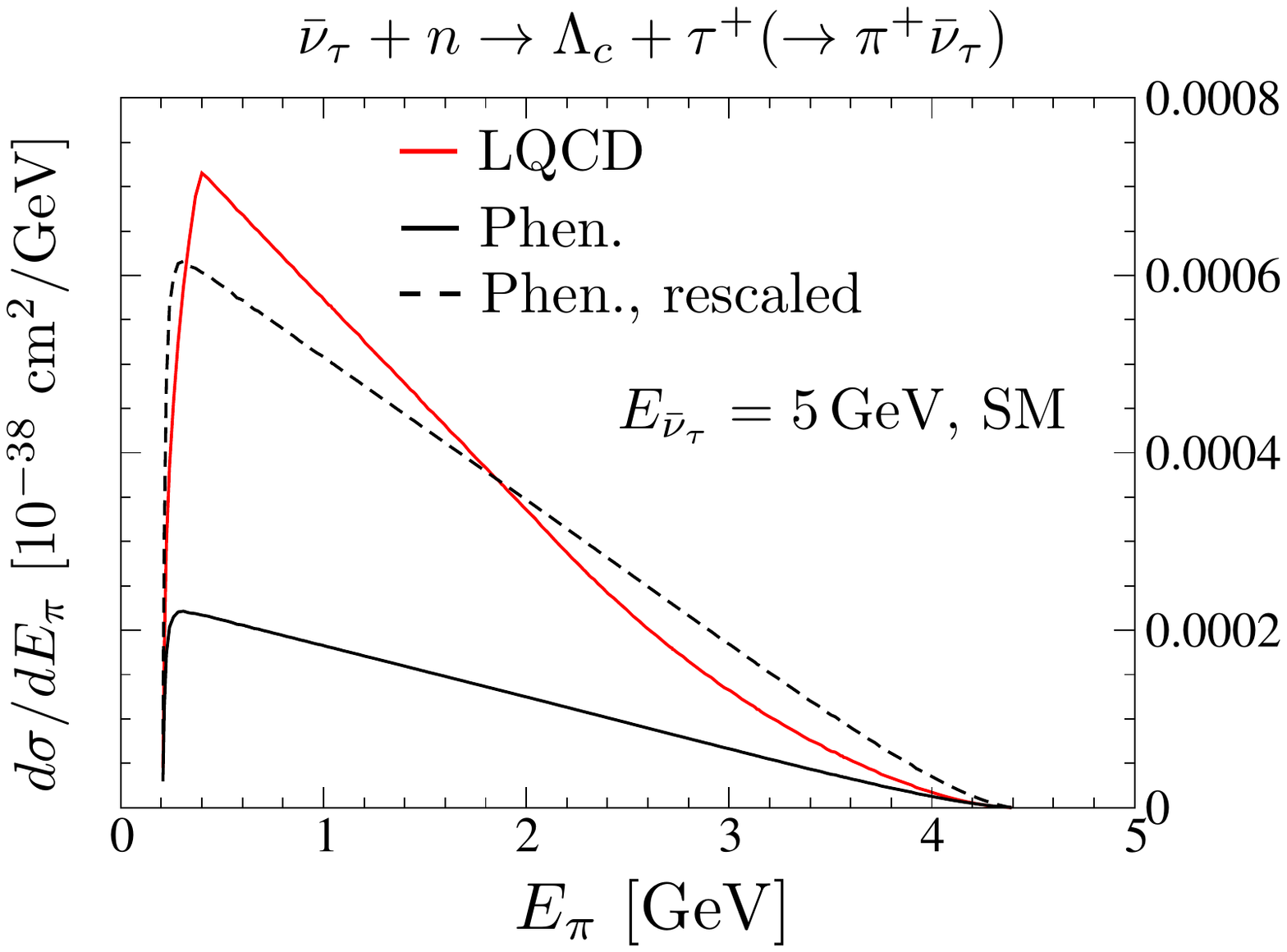}}
\rotatebox{270}{\includegraphics[height=.45\textwidth]{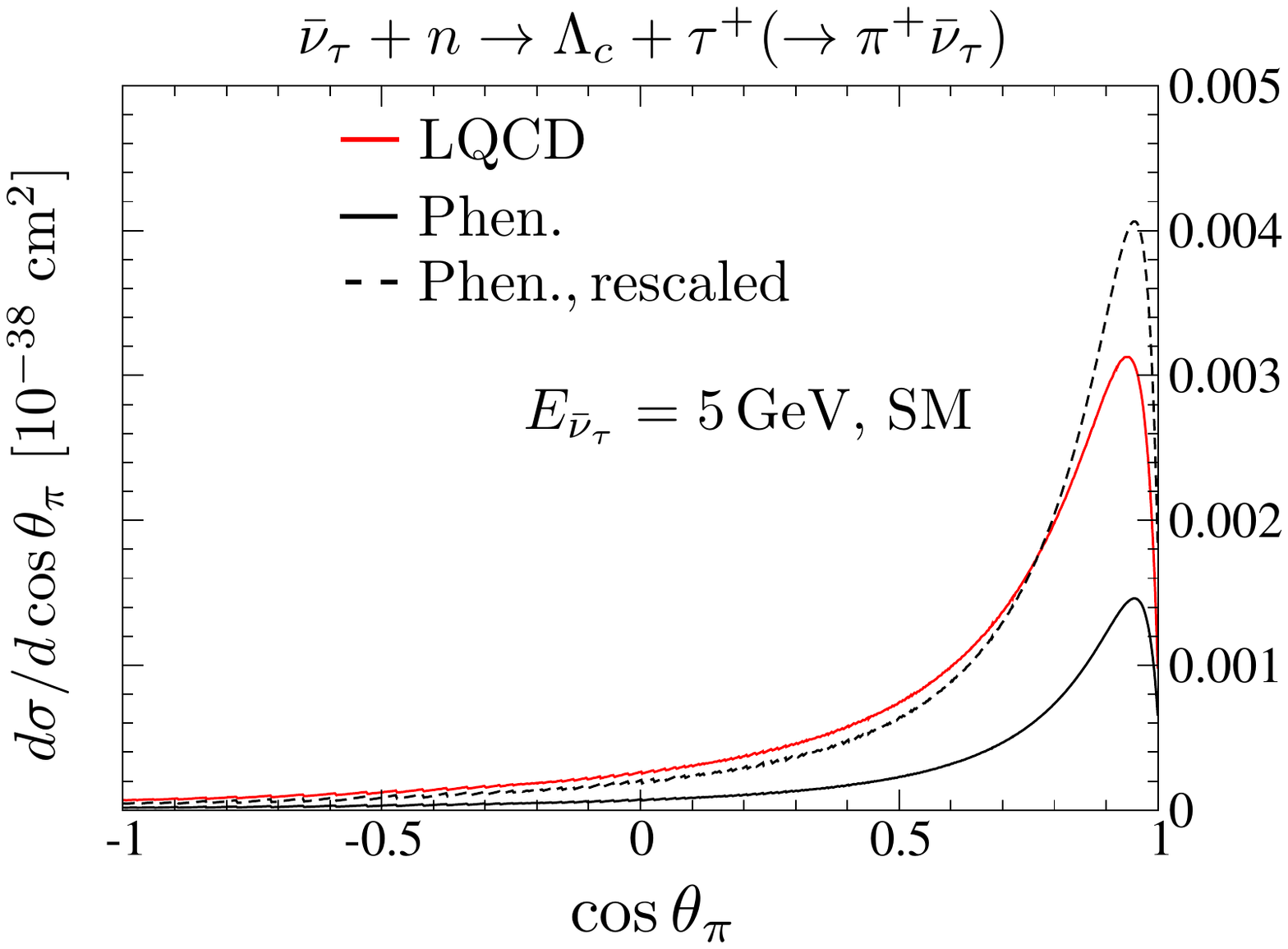}}\caption{SM differential cross sections $d\sigma/dE_\pi$ (left panel) and
$d\sigma/d\cos\theta_\pi$ (right panel) for the
$\bar\nu_\tau p \to \Lambda \tau^+ (\pi^+ \bar\nu_\tau)$ reaction,
evaluated at $E_{\bar\nu_\tau} = 5\,\mathrm{GeV}$ for a fully right-handed
polarized $\bar\nu_\tau$ beam. Results obtained using the LQCD form factors from Ref.~\cite{Bacchio:2025auj}
(red curve) and the phenomenological form factors from
Ref.~\cite{Mintz:2004hm} (black curve) are shown. The dotted curve
corresponds to the latter prediction rescaled to reproduce the total
cross section obtained with the LQCD form factors.}  
\label{fig:dsigdepilambda}
\end{center}
\end{figure}

In Fig.~\ref{fig:dsigdepilambda} we present the SM single-differential cross
sections $d\sigma/dE_\pi$ (left panel) and $d\sigma/d\cos\theta_\pi$
(right panel). The angular distribution exhibits a clear forward
peaking, although less pronounced than in the
$\nu_\tau n \to \Lambda_c \tau^- (\pi^- \nu_\tau)$ reaction, while the
energy distribution is dominated by the low-energy region of the pion
spectrum. The differences in the overall normalization obtained with the
two form-factor sets are now clearly visible. In addition, noticeable
shape differences are observed: the distributions computed with the
phenomenological form factors of Ref.~\cite{Mintz:2004hm} peak at larger
values of $\cos\theta_\pi$ (closer to 1) and at lower pion energies $E_\pi$.

In Table~\ref{tab:asylambda}, we present the results for the CP asymmetry $\Delta\sigma$, 
defined in Eq.~\eqref{eq:sig2}. In this case, tensor form factors are not available, 
so we focus on the effects of nonzero, purely imaginary $C^S_{su\tau LL}$ or 
$C^S_{su\tau RL}$ Wilson coefficients. We assume that only one of these coefficients 
is nonzero at a time, with its magnitude kept moderate, not exceeding 0.1. 
The resulting asymmetry depends on the choice of form-factor set, and the values should 
be regarded as indicative of the expected order of magnitude in the presence of 
CP-violating new physics.

\begin{table}[tbh]
\begin{center}
\begin{tabular}{|c|c|c|c|c|}\hline
&\multicolumn{2}{c|}{FFs from Ref.~\cite{Bacchio:2025auj}}&\multicolumn{2}{c|}{FFs from Ref.~\cite{Mintz:2004hm}}\\\hline
$C^S_{su\tau LL}$      &  $\frac{\Delta\sigma}{\sigma}(\%)$ & 
      $\frac{\sigma-\sigma_{\rm SM}}{\sigma_{\rm SM}}(\%)$&  $\frac{\Delta\sigma}{\sigma}(\%)$ & 
      $\frac{\sigma-\sigma_{\rm SM}}{\sigma_{\rm SM}}(\%)$\\\hline
$0.01i$&$-0.26$& 0.51&$0.13$&0.01\\ 
$0.1i$&$-1.74$&51&$1.26$&1.08\\\hline 
\end{tabular}
\begin{tabular}{|c|c|c|c|c|}\hline
&\multicolumn{2}{c|}{FFs from Ref.~\cite{Bacchio:2025auj}}&\multicolumn{2}{c|}{FFs from Ref.~\cite{Mintz:2004hm}}\\\hline
$C^S_{su\tau RL}$      &  $\frac{\Delta\sigma}{\sigma}(\%)$ & 
      $\frac{\sigma-\sigma_{\rm SM}}{\sigma_{\rm SM}}(\%)$&  $\frac{\Delta\sigma}{\sigma}(\%)$ & 
      $\frac{\sigma-\sigma_{\rm SM}}{\sigma_{\rm SM}}(\%)$\\\hline
$0.01i$&$0.40$& 0.51&$0.14$&0.01\\ 
$0.1i$&$2.68$&51&$1.4$&1.08\\\hline 
\end{tabular}
\end{center}
\caption{CP asymmetry $\Delta\sigma$ [Eq.~\eqref{eq:sig2}] and variations of the total cross section
($\sigma - \sigma_{\rm SM}$) for the
$\bar\nu_\tau p \to \Lambda \tau^+ (\pi^+ \bar\nu_\tau)$ reaction, evaluated at
$E_{\bar\nu} = 5\,\mathrm{GeV}$ for a fully right-handed $\bar\nu_\tau$ beam.
The NP effects are generated by assuming that one of the scalar Wilson coefficients
$C^S_{su\tau LL}$ or $C^S_{su\tau RL}$ is nonzero and purely imaginary.
Two sets of hadronic form factors are considered: the LQCD form factors from
Ref.~\cite{Bacchio:2025auj} and the phenomenological form factors from
Ref.~\cite{Mintz:2004hm}, which yield
SM total cross sections of $\sigma_{\rm SM} = 0.00124\times10^{-38}\,$cm$^2$ and $\sigma_{\rm SM} = 0.00045\times10^{-38}\,$cm$^2$, respectively.  
See the text for further details.}
\label{tab:asylambda}
\end{table} 

Examining the change in the cross section relative to the SM value, also shown in Table~\ref{tab:asylambda}, 
we observe a large increase when using the LQCD form factors with a Wilson coefficient modulus of 0.1. 
This result is somewhat unexpected given the small values considered for the NP Wilson coefficients 
and points to issues with the extrapolation of some form factors to large negative $q^2$ values.
The LQCD form factors from Ref.~\cite{Bacchio:2025auj} were determined in the range 
$q^2 \in [-0.25\,\mathrm{GeV}^2, q^2_\mathrm{max}]$, with $q^2_\mathrm{max} = (M_\Lambda - M_p)^2 \approx 0.0315~\mathrm{GeV}^2$. 
However, for the present reaction at $E_{\bar\nu_\tau} = 5~\mathrm{GeV}$, values as negative as $-5~\mathrm{GeV}^2$ can be reached. 
This makes the extrapolation to this $q^2$ region somewhat problematic.
We attribute the large change in the cross section including NP effects to the axial helicity form factor $g_0$. 
This does not imply that $g_0$ is improperly evaluated in Ref.~\cite{Bacchio:2025auj} within the range 
$q^2 \in [-0.25\,\mathrm{GeV}^2, q^2_\mathrm{max}]$; rather, it indicates that extrapolating the provided 
parametrization to large negative $q^2$ values is likely not justified. 
The parametrizations in Ref.~\cite{Bacchio:2025auj} do not include explicit poles at positive $q^2$, 
causing some form factors to approach zero slowly as $q^2 \to -\infty$, 
in contrast to the phenomenological form factors from Ref.~\cite{Mintz:2004hm}.

\section{Conclusions}
\label{sec:concl}

We have derived general expressions for $\tau$ (anti-)neutrino--induced production of strange and charmed
hyperons on nucleons, incorporating all dimension-six BSM operators contributing to the charged-current
transitions $\bar\nu_\tau u \to s \tau^+$ and $\nu_\tau d \to c \tau^-$, and allowing for both left- and
right-handed neutrino fields. The final-state $\tau$ lepton does not propagate far enough to produce a
displaced vertex, and its decay necessarily involves at least one additional neutrino. As a consequence,
extracting complete information from the underlying weak vertex—ideally the full set of components of the
polarization four-vector over the two-dimensional kinematical domain~\cite{Penalva:2021gef}—requires exploiting
the visible kinematics of the decay products in the subsequent $\tau$ decay.
In this work we focused on $\nu_\tau n \to \Lambda_c \tau^- (\pi^- \nu_\tau)$ and $\bar\nu_\tau p \to \Lambda \tau^+ (\pi^+ \bar\nu_\tau)$ sequential reactions. This allowed us to construct a CP asymmetry observable, $\Delta\sigma$, by separately counting events in which the pion is emitted above or below the plane defined by the incoming neutrino and the outgoing hyperon. A nonzero value would constitute a clean and unambiguous signature of CP-violating new physics beyond the SM.

We have obtained estimates of the SM and NP pion single- and double-differential cross sections,
as well as of the total integrated cross sections and  $\Delta\sigma$, for both $\Lambda_c$ and $\Lambda$ hyperon production. Following the strategy of Ref.~\cite{Penalva:2022vxy}, the statistical sensitivity of the pion energy and angular distributions is enhanced by integrating over the outgoing hyperon energy.
For NP predictions we adopted a conservative approach and considered moderate deviations
of order $\mathcal{O}(10^{-1})$ in the relevant Wilson coefficients.
The resulting effects are generally small, yet potentially observable. The corresponding CP-asymmetry ratios $\Delta\sigma/\sigma$ are  expected to reach, at most, the percent level. Depending on the complex phase of the Wilson coefficient under consideration, they may either enhance or suppress the differential cross sections while largely preserving the overall shapes of the distributions.  This complicates the extraction of a clear NP signal due to the inherent uncertainties in the SM calculation.
The dominant source of theoretical uncertainty originates from the limited knowledge of the hadronic form factors in the $q^2 < 0$ region relevant for (anti-)neutrino--nucleon scattering. 
Those uncertainties imply that the results presented in this work should be interpreted as approximate estimates. 
In order to reduce the sensitivity to hadronic form factors, we introduced the CP-odd asymmetry $\Delta \sigma$ defined as a ratio of partially integrated cross sections. Furthermore, any nonzero value of this asymmetry would constitute a clear signal of CP-violating new physics beyond the SM,  irrespective of the specific form-factor inputs.

In future studies of these reactions in a nucleus, the $\tau$ momentum will be no longer uniquely determined by the momenta  of the incoming neutrino and the outgoing hyperon. Consequently, in the absence of direct  $\tau$-lepton detection---which is experimentally challenging---the lepton ($\nu_\tau \tau^-$/$\bar\nu_\tau \tau^+$) plane cannot be defined. As a result, although the asymmetry in Eq.~\eqref{eq:sig2} can still be considered when final-state interactions of the outgoing hyperon are neglected, it would no longer depend uniquely on the CP-odd quantity ${\cal P}_{TT}$.

In conclusion, the formalism developed in this work is fully general and readily applicable to weak 
transitions involving other quark flavors---such as neutron--proton or proton--neutron 
processes---as well as to the production of different hyperons. Requiring only the relevant 
hadronic form factors and NP Wilson coefficients, it provides a flexible and robust framework for 
exploring NP in a broad class of neutrino weak processes. In principle, such reactions could be 
investigated at the nucleus level at the DUNE far detector, where ``a modest but unique $\nu_\tau$ data sample'' \cite{DeGouvea:2019kea} is expected.

\section*{Acknowledgments}
This work has been partially supported by the Spanish 
MICIU/AEI/10.13039/501100011033 under 
grants   PID2022-141910NB-I00 and PID2023-147458NB-C21, by 
Generalitat Valenciana's PROMETEO
grant CIPROM/2023/59, by the “Planes Complementarios de I+D+i” program
(Grant No. ASFAE/2022/022) from MICINN with funding from
the European Union NextGenerationEU and Generalitat
Valenciana, and by the JCyL grant SA091P24 under program 
EDU/841/2024.

\appendix
\section{Lepton Tensor, $\tau$ Polarization Vector, and Phase-Space Integration in $\nu_\tau n \to \Lambda_c \tau^- (\pi^- \nu_\tau)$}
\label{app:leptonANDPandKin}

\subsection{Lepton Tensor}
\label{app:lepton}
The tensor for the two neutrinos and the final pion is given by the product of lepton currents, ${\cal L}_{\chi ab} = l_{\chi a}l^\dagger_{\chi b}$, with
\bea
l_{\chi a}&=&2\,G_F f_\pi V^*_{ud}\,\bar u_{\nu_\tau}(\vec p_{\nu_\tau},L)
\slashed{p}_\pi P_L\frac{\slashed{k}' +m_\tau}{k^{\prime2}-m^2_\tau+i
\sqrt{k^{\prime2}}\,\Gamma_\tau(\sqrt{k^{\prime 2}}\,)}
\Gamma_a \frac1{\sqrt2}P_\chi u_{\nu_\tau }(\vec k,\chi),
\eea
where $\Gamma_a = \gamma_\mu, I $ and $ \sigma_{\mu\nu}$, and $u$ is a dimensionful Dirac spinor. We have employed the effective $\tau\to\pi\nu_\tau$ vertex
\be
{\cal L}_{\tau\to\pi\nu_\tau}=-i2G_F f_\pi V^*_{ud}\bar
\Psi_{\nu_\tau}(x) \gamma^\mu\ P_L\Psi_\tau(x)
\partial_\mu \Phi_\pi(x),
\ee
where $f_\pi \approx 93$ MeV is the pion decay constant, $V_{ud}$ the CKM matrix element for $W^- \to \bar u d$, $k'$ the intermediate $\tau$ four-momentum, and $\Gamma_\tau$ its total width, dependent on the invariant mass; then, using $p_\pi = k' - p_{\nu_\tau}$ and assuming massless neutrinos, $l_{\chi a}$ can be rewritten as
\bea
l_{\chi a }&=&2\,G_F f_\pi V^*_{ud}\,\bar u_{\nu_\tau}(\vec p_{\nu_\tau},L)
 P_R\frac{k^{\prime2} +m_\tau\slashed{k}'}{k^{\prime2}-m^2_\tau+i
\sqrt{k^{\prime2}}\,\Gamma_\tau(\sqrt{k^{\prime 2}}\,)}
\Gamma_a \frac1{\sqrt2}P_\chi u_{\nu_\tau }(\vec k,\chi).
\eea
Since $\Gamma_\tau(\sqrt{k^{\prime 2}}\,)/m_\tau\ll1$, one can 
approximate 
\bea
\frac{1}{(k^{\prime2}-m^2_\tau)^2+
k^{\prime2}\Gamma^2_\tau(\sqrt{k^{\prime 2}}\,)}\approx\frac{\pi}
{m_\tau\Gamma_\tau(m_\tau)}\delta(k^{\prime2}-m^2_\tau),
\eea
an from there one obtains\footnote{The presence of the chirality projectors in $l_{\chi a} l^\dagger_{\chi b}$ permits summation over the chiralities of both neutrinos, thereby reducing the calculation to a Dirac trace.}
\bea
{\cal L}_{\chi ab}= l_{\chi a} l^{\dagger}_{\chi b}=
\frac{4\pi m_\tau G_F^2f_\pi^2|V_{ud}|^2}{\Gamma_\tau(m_\tau)}\,{\rm Tr}\,[\slashed{p}_{\nu_\tau} P_R(\slashed{k}'+m_\tau)
\Gamma_a \frac1{2}P_\chi\slashed{k}\gamma^0\Gamma^\dagger_b\gamma^0
(\slashed{k}'+m_\tau)P_L]\,\delta(k^{\prime2}-m^2_\tau),
\eea
Further using 
\bea
&&\Gamma_{\tau\to\pi\nu_\tau}(m_\tau)=
G_F^2f_\pi^2|V_{ud}|^2
\frac{(m^2_\tau-m^2_\pi)^2}{8\pi m_\tau},
\eea
we arrive at Eq.~\eqref{eq:calL} and to
\bea
\sigma&=&{\cal B}(\tau\to\pi\nu_\tau)\frac{m^2_\tau M_{\Lambda_c}
G_F^2|V_{cd}|^2}
{2\pi^3|\vec k\,|(m^2_\tau-m^2_\pi)^2}
\int\frac{d^3p'}{E'}\int\frac{d^3p_\pi}{E_\pi }\int\frac{d^3p_{\nu_\tau}}
{ E_{\nu_\tau}}
\delta^{(4)}(k+p-p'-p_\pi-p_{\nu_\tau})
\nonumber\\
&\times&\delta((k+p-p')^2-m^2_\tau)\sum_{\chi=L,R}\rho_{\chi\chi}\sum_{a,b}W^{ab}_\chi\ {\rm Tr}\,[\slashed{p}_{\nu_\tau} P_R(\slashed{k}'+m_\tau)
\Gamma_a \frac1{2}P_\chi\slashed{k}\gamma^0\Gamma^\dagger_b\gamma^0
(\slashed{k}'+m_\tau)P_L],
\eea
Using now that $p_{\nu_\tau}=k'-p_\pi$, we have
\bea
&&{\rm Tr}\,[\slashed{p}_{\nu_\tau} P_R(\slashed{k}'+m_\tau)
\Gamma_a \frac1{2}P_\chi\slashed{k}\gamma^0\Gamma^\dagger_b\gamma^0
(\slashed{k}'+m_\tau)P_L]\nonumber\\
&&=(m_\tau^2-p_\pi\cdot k') {\rm Tr}\,[(\slashed{k}'+m_\tau)
\Gamma_a \frac12P_\chi\slashed{k}\gamma^0\Gamma^\dagger_b\gamma^0]
-{\rm Tr}\,[\slashed{p}_\pi \frac12\gamma_5(\slashed{k}'+m_\tau)
\Gamma_a \frac1{2}P_\chi\slashed{k}\gamma^0\Gamma^\dagger_b\gamma^0
(\slashed{k}'+m_\tau)]\nonumber\\
&&=\frac{m_\tau^2-m_\pi^2}2 {\rm Tr}\,[(\slashed{k}'+m_\tau)
\Gamma_a \frac12P_\chi\slashed{k}\gamma^0\Gamma^\dagger_b\gamma^0]
+\frac12{\rm Tr}\,[(\slashed{k}'+m_\tau)
\Gamma_a \frac1{2}P_\chi\slashed{k}\gamma^0\Gamma^\dagger_b\gamma^0
(\slashed{k}'+m_\tau)\gamma_5\slashed{p}_\pi].\label{eq:appearP}
\eea
which then leads to Eq.~\eqref{eq:sec1}.
\subsection{Polarization vector}
\label{app:P}

We provide here the details of the $\tau$ polarization vector introduced in Eq.~\eqref{eq:P}. In Ref.~\cite{Hernandez:2025snr}, it was shown that
\bea
\widetilde{\cal N}(\omega,k\cdot p)=\frac{1}{M^2_n}\,\sum_{\chi=L,R}\rho_{\chi\chi}\sum_{a,b}W^{ab}_\chi L_{\chi ab}=
\frac{1}{2}\Big[-{\cal A}(\omega)+{\cal B}(\omega)
\frac{p\cdot k}{M^2_n}-{\cal C}(\omega)
\frac{(p\cdot k)^2}{M^4_n}\Big], \label{eq:nopol}
\eea
Proceeding as in Ref.~\cite{Penalva:2021gef}, the polarization four-vector ${\cal P}^\mu$ reads
\bea
{\cal P}^\mu &=&\frac1{\widetilde{\cal N}(\omega,k\cdot p)}\Big[
\frac{p^\mu_\perp}{M_n}\widetilde{\cal N}_{{\cal H}1}(\omega,k\cdot p)+
\frac{q^\mu_\perp}{M_n}\widetilde{\cal N}_{{\cal H}2}(\omega,k\cdot p)+
\frac{\epsilon^{\mu k' q p}}{M^3_n}\widetilde{\cal N}_{{\cal H}3}
(\omega,k\cdot p)\Big],\\
\widetilde{\cal N}_{{\cal H}1}(\omega,k\cdot p)&=&-{\cal A}_{{\cal
H}}(\omega)+{\cal C}_{{\cal
H}}(\omega)
\frac{p\cdot k}{M^2_n} \label{eq:pol1},\\
\widetilde{\cal N}_{{\cal H}2}(\omega,k\cdot p)&=&-{\cal B}_{{\cal
H}}(\omega)+{\cal D}_{{\cal
H}}(\omega)
\frac{p\cdot k}{M^2_n}-{\cal E}_{{\cal
H}}(\omega)
\frac{(p\cdot k)^2}{M^4_n},\label{eq:pol2}\\
\widetilde{\cal N}_{{\cal H}3}(\omega,k\cdot p)&=&-{\cal F}_{{\cal
H}}(\omega)+{\cal G}_{{\cal
H}}(\omega)
\frac{p\cdot k}{M^2_n}.\label{eq:pol3}
\eea
The expressions for the functions ${\cal A}(\omega)$, ${\cal B}(\omega)$, and ${\cal C}(\omega)$ in Eq.~\eqref{eq:nopol}, and for ${\cal A}_{\cal H}(\omega)$, ${\cal B}_{\cal H}(\omega)$, ${\cal C}_{\cal H}(\omega)$, ${\cal D}_{\cal H}(\omega)$, ${\cal E}_{\cal H}(\omega)$, ${\cal F}_{\cal H}(\omega)$, and ${\cal G}_{\cal H}(\omega)$ in Eqs.~\eqref{eq:pol1}–\eqref{eq:pol3}, in terms of the $\widetilde W_\chi$ structure functions are given in Appendix~D of Ref.~\cite{Penalva:2021wye}. In that reference, the quantities $M$, $M_\omega$, and $m_\ell$ correspond to $M_n$, $(M_n - M_{\Lambda_c}\omega)$, and $m_\tau$, respectively. There, these functions are obtained by summing over $\chi = L, R$, as the analysis involved a final-state neutrino with a small but non-vanishing mass, allowing both chiralities, while neutrino-mass–suppressed and left–right interference terms were neglected. In the present case, the neutrino is incoming, and for a purely left-handed neutrino (right-handed antineutrino) beam produced by SM interactions, only the $\chi = L$ ($\chi = R$) contributions should be retained. More generally, a factor $\rho_{\chi\chi}$ must be included in the chirality sums entering the definition of these functions.

Eqs.~\eqref{eq:nopol} and \eqref{eq:pol1}–\eqref{eq:pol3} follow directly from crossing symmetry, with the scattering reaction differing from the decay in Ref.~\cite{Penalva:2021wye} by the interchange of the neutrino four-momentum $k$ between initial and final states.

The longitudinal and transverse components of the polarization vector in Eq.~\eqref{eq:plttt} are~\cite{Penalva:2021gef}
\bea
{\cal P}_{L}&=&-\frac1{\widetilde{\cal N}(\omega,k\cdot p)}\Big[
\frac{|\vec k\,'|}{m_\tau}\widetilde{\cal N}_{{\cal H}1}(\omega,k\cdot p)-
\frac{|\vec k\,|}{M_n m_\tau}\Big[|\vec k\,'|-\frac{k^{\prime0}}{|\vec k\,'|}(|\vec k\,|-
|\vec p\,'|\cos\theta')\Big]\widetilde{\cal N}_{{\cal H}2}(\omega,k\cdot p)\Big],\no
{\cal P}_{T}&=&
\frac{|\vec k\,|\,|\vec p\,'|\sin\theta'}{|\vec k\,'|M_n }
\frac{\widetilde{\cal N}_{{\cal H}2}(\omega,k\cdot p)}{\widetilde{\cal N}(\omega,k\cdot p)},
\no
{\cal P}_{TT}&=&
\frac{|\vec k\,|\,|\vec p\,'|\sin\theta'}{M^2_n}
\frac{\widetilde{\cal N}_{{\cal H}3}(\omega,k\cdot p)}{\widetilde{\cal N}(\omega,k\cdot p)},
\eea
where we have taken the $Z^+$ axis along the incident neutrino three-momentum $\vec k$ and where
\bea
|\vec k\,'|=\sqrt{|\vec p\,'|^2+|\vec k\,|^2-2|\vec p\,'|\,
|\vec k\,|\cos\theta'}\ ,\ k^{\prime0}=\sqrt{|\vec k\,'|^2+m^2_\tau}.
\eea
are the three-momentum and energy of the intermediate tau lepton.

\subsection{Phase-Space Integration }
\label{app:kin}

Starting from Eq.~\eqref{eq:sec1} for the cross section, for each value of $\vec p\,'$ we can now take a rotation $R$ around the 
 $Z$ axis, defined by $\vec k$,    such that  $\tilde p'=R p\,'=(E',-|\vec
 p\,'|\sin\theta',0,|\vec
 p\,'|\cos\theta')$. Equivalently, on an event-by-event basis, the $Y^+$ axis is defined along $\vec p\,' \times \vec k$, while the $X^+$ axis lies along $(\vec{p}\,' \times \vec k) \times \vec k$, as shown in Fig.~\ref{fig:kin}. Using this transformation, Lorentz invariance, and the changes of variables $R\vec p_\pi \to \vec p_\pi$ and $R\vec p_{\nu_\tau} \to \vec p_{\nu_\tau}$ (with $R p = p$ and $R k = k$), the integration over the $\Lambda_c$ azimuthal angle becomes trivial, leading to
\bea
\sigma
&=&{\cal B}(\tau\to\pi\nu_\tau)\frac{ m^2_\tau M_{\Lambda_c}G_F^2|V_{cd}|^2}
{2\pi^2|\vec k\,|(m^2_\tau-m^2_\pi)}
\int d\cos\theta'\,|\vec p\,'| dE'\int\frac{d^3p_\pi}{ E_\pi}\int
\frac{d^3p_{\nu_\tau}}{ E_{\nu_\tau}}
\delta^{(4)}(k+p-\tilde p'-p_\pi-p_{\nu_\tau})
\nonumber\\
&\times&\delta((k+p-\tilde p')^2-m^2_\tau)\Big[\sum_{\chi=L,R}\rho_{\chi\chi}\sum_{a,b}W^{ab}_\chi L_{\chi ab}\Big]
\Big(1+\frac{2m_\tau}{m^2_\tau-m^2_\pi}\,p_\pi\cdot
 R{\cal P}
\Big),
\eea
with
\bea
 R{\cal P}&=&{\cal P}_L\, RN_L+{\cal P}_T\,  RN_T+
{\cal P}_{TT}\,  RN_{TT},
\no
(RN_L)^\mu&=&(\frac{|\vec
k\,'|}{m_\tau},\frac{k^{\prime0}}{m_\tau|\vec k\,'|}|\vec
p\,'|\sin\theta',0,\frac{k^{\prime0}}{m_\tau|\vec k\,'|}(|\vec k\,|-
|\vec p\,'|\cos\theta')\,),\nonumber\\
(RN_T)^\mu&=&\frac1{|\vec k\,'|}
(0,|\vec k\,|-|\vec p\,'|\cos\theta',0,-|\vec
p\,'|\sin\theta'),\nonumber\\
(RN_{TT})^\mu&=&(0,0,1,0),
\eea
and 
\bea
p_\pi\cdot R{\cal P}&=&\frac{{\cal P}_L}{m_\tau}\Big[
E_\pi|\vec k\,'|-\frac{|\vec p_\pi|k^{\prime0}}{|\vec k\,'|}
(|\vec p\,'|\sin\theta_\pi\cos\varphi_\pi\sin\theta'+
\cos\theta_\pi(|\vec k\,|-|\vec p\,'|\cos\theta')\,)\Big]\no
&+&\frac{{\cal P}_T|\vec p_\pi|}{|\vec k\,'|}\Big[
-(|\vec k\,|-|\vec p\,'|\cos\theta')\sin\theta_\pi\cos\varphi_\pi+
|\vec p\,'|\sin\theta'\cos\theta_\pi\Big]-{\cal P}_{TT}\,|\vec p_\pi|\sin\theta_\pi\sin\varphi_\pi.
\eea
We now introduce $\theta_\tau$, the polar angle of the virtual $\tau$, and $\cos\theta_{\tau\pi}$, the cosine of the angle between the three-momenta of the $\tau^-$ and the $\pi^-$ (see Fig.~\ref{fig:kin}),
\bea
\vec k\,'&=&(|\vec p\,'|\sin\theta',0,|\vec k\,|-|\vec p\,'|\cos\theta')=
|\vec k\,'|(\sin\theta_\tau,0,\cos\theta_\tau),\\
\cos\theta_{\tau\pi}&=&\sin\theta_\tau\sin\theta_\pi\cos\varphi_\pi
+\cos\theta_\tau\cos\theta_\pi,
\eea
and rewrite
\bea
p_\pi\cdot R{\cal P}&=&\frac{{\cal P}_L}{m_\tau}[
E_\pi|\vec k\,'|-|\vec p_\pi|k^{\prime0}\cos\theta_{\tau\pi}]+\frac{{\cal P}_T|\vec p_\pi|}{\sin\theta_\tau}[
\cos\theta_\pi-\cos\theta_\tau\cos\theta_{\tau\pi}]
-{\cal P}_{TT}|\vec p_\pi|\sin\theta_\pi\sin\varphi_\pi,
\eea
which leads to 
\bea
\sigma
&=&{\cal B}(\tau\to\pi\nu_\tau)\frac{ m^2_\tau M_{\Lambda_c}G_F^2|V_{cd}|^2}
{2\pi^2|\vec k\,|(m^2_\tau-m^2_\pi)}
\int d\cos\theta'\,|\vec p\,'| dE'\int\frac{d^3p_\pi}{ E_\pi}\int
\frac{d^3p_{\nu_\tau}}{ E_{\nu_\tau}}
\delta^{(4)}(k+p-\tilde p'-p_\pi-p_{\nu_\tau})
\nonumber\\
&\times&\delta((k+p-\tilde p')^2-m^2_\tau)\Big[\sum_{\chi=L,R}\rho_{\chi\chi}\sum_{a,b}W^{ab}_\chi L_{\chi ab}\Big]
\Bigg\{1+\frac{2m_\tau}{m^2_\tau-m^2_\pi}\,\Big({\cal P}_L\frac1{m_\tau}[
E_\pi|\vec k\,'|-|\vec p_\pi|k^{\prime0}\cos\theta_{\tau\pi}]\no
&+&\frac{{\cal P}_T|\vec p_\pi|}{\sin\theta_\tau}[
\cos\theta_\pi-\cos\theta_\tau\cos\theta_{\tau\pi}]-{\cal P}_{TT}|\vec p_\pi|\sin\theta_\pi\sin\varphi_\pi\Big)
\Bigg\}, \label{eq:pttANDsinphipi}
\eea
Now using the delta functions to integrate over $d^3 p_{\nu}$, $d\cos\theta'$ and $d\varphi_\pi$, 
we obtain 
\bea
\sigma
&=&{\cal B}(\tau\to\pi\nu_\tau)\frac{ m^2_\tau M_{\Lambda_c}G_F^2|V_{cd}|^2}
{2\pi^2|\vec k\,|^2(m^2_\tau-m^2_\pi)}
\int \frac{H(1-|\cos\theta'_0|)}
{\sqrt{(|\vec k\,|+M_n-E')^2-m^2_\tau}}dE'\int
d\cos_\pi dE_\pi
\frac{H(1-|\cos\varphi^0_\pi|)}{\sin\theta^0_\tau\sin\theta_\pi|\sin\varphi^0_\pi|}
\nonumber\\
&\times&\Big[\sum_{\chi=L,R}\rho_{\chi\chi}\sum_{a,b}W^{ab}_\chi L_{\chi ab}\Big]
\Bigg\{1+
\frac{2m_\tau}{m^2_\tau-m^2_\pi}\,\Big(\frac{{\cal P}_L}{m_\tau}[
E_\pi\sqrt{|\vec k\,|^2+|\vec p\,'|^2-2|\vec k\,|\,|\vec p\,'|\cos\theta'_0}\no
&-&|\vec p_\pi|(|\vec k\,|+M_n-E')\cos\theta^0_{\tau\pi}]+\frac{{\cal P}_T|\vec p_\pi|}{\sin\theta^0_\tau}[
\cos\theta_\pi-\cos\theta^0_\tau\cos\theta^0_{\tau\pi}]\Big)
\Bigg\},\label{eq:secc-diff}
\eea
where $\cos\theta'_0$, $\cos\theta^0_{\tau}$, $\sin\theta^0_{\tau}$, $\cos\varphi^0_\pi$ and $\cos\theta^0_{\tau\pi}$ are fixed by the momentum-conservation delta functions and given by
\bea
\cos\theta'_0&=&\frac{2E'(|\vec k\,|+M_n)+m^2_\tau-(M^2_n+2M_n|\vec
k\,|+M^2_{\Lambda_c})}{2|\vec k\,|\,|\vec p\,'|},\no
\cos\theta^0_{\tau}&=&\frac{|\vec k\,|-|\vec p\,'|\cos\theta'_0}
{\sqrt{(|\vec k\,|+M_n-E')^2-m^2_\tau}}\ ,\ \sin\theta^0_{\tau}=\frac{|\vec p\,'|\sin\theta'_0}
{\sqrt{(|\vec k\,|+M_n-E')^2-m^2_\tau}},\no
\cos\varphi^0_\pi&=&\frac{2(|\vec k\,|+M_n-E')E_\pi-m^2_\tau-m^2_\pi-2|\vec
p_\pi|\sqrt{(|\vec k\,|+M_n-E')^2-m^2_\tau}\cos\theta^0_\tau\cos\theta_\pi}{
2|\vec p_\pi|\sqrt{(|\vec k\,|+M_n-E')^2-m^2_\tau}\sin\theta^0_\tau\sin\theta_\pi},\no
\cos\theta^0_{\tau\pi}&=&\sin\theta^0_\tau\sin\theta_\pi\cos\varphi^0_\pi
+\cos\theta^0_\tau\cos\theta_\pi=\frac{2(|\vec k\,|+M_n-E')E_\pi-m^2_\tau-m^2_\pi}
{2|\vec p_\pi|\sqrt{(|\vec k\,|+M_n-E')^2-m^2_\tau}}.\label{eq:aux1}
\eea
After integrating over $\varphi_\pi$, Eq.~\eqref{eq:aux1} no longer retains sensitivity to ${\cal P}_{TT}$. However, an angular asymmetry in the pion azimuthal angle can be constructed,
\bea
\Delta\sigma &=&\sigma(0<\varphi_\pi<\pi)-
\sigma(\pi<\varphi_\pi<2\pi)\no
&=&{\cal B}(\tau\to\pi\nu_\tau)\frac{ m^2_\tau M_{\Lambda_c}G_F^2|V_{cd}|^2}
{2\pi^2|\vec k\,|^2(m^2_\tau-m^2_\pi)}
\int \frac{H(1-|\cos\theta'_0|)}
{\sqrt{(|\vec k\,|+M_n-E')^2-m^2_\tau}}dE'\int
d\cos_\pi dE_\pi
\frac{H(1-|\cos\varphi^0_\pi|)}{\sin\theta^0_\tau\sin\theta_\pi|\sin\varphi^0_\pi|}
\nonumber\\
&&\hspace{1cm}\times\Big[\sum_{\chi=L,R}\rho_{\chi\chi}\sum_{a,b}W^{ab}_\chi L_{\chi ab}\Big]
\,\frac{2m_\tau}{m^2_\tau-m^2_\pi}\Big[-{\cal P}_{TT}\, |\vec p_\pi|\,
|\sin\varphi^0_\pi|\,\sin\theta_\pi\Big],
\label{eq:asim}
\eea
which probes ${\cal P}_{TT}$ and thus CP-violating new physics beyond the SM.

The Heaviside function $H(1-|\cos\theta'_0|)$ in Eqs.~\eqref{eq:secc-diff} and \eqref{eq:asim} constrains $E'$ to the interval $E' \in [E'_-, E'_+]$, where $E'_\pm$ correspond to the maximum and minimum energies available to the final $\Lambda_c$ when produced in association with a $\tau$ lepton,
\bea
\ E'_\pm=(|\vec k\,|+M_n) 
\frac{S+M^2_{\Lambda_c}-m^2_\tau}{2S}\pm
|\vec k\,|\frac{\lambda^{1/2}(S, M^2_{\Lambda_c},m^2_\tau)}{2S},\label{eq:Eprimelimits} 
\eea
with $\sqrt{S}=\sqrt{M^2_n+2|\vec k\,|\,M_n}$ the total center of mass energy and
$\lambda(a,b,c)=a^2+b^2+c^2-2ab-2ac-2bc$ the K\"allen lambda function.

For a given $E'$, the condition $|\cos\theta^0_{\tau\pi}|\le 1$ restricts the pion energy to the interval $E_\pi \in [E_\pi^-(E'), E_\pi^+(E')]$.
\bea
E^{\pm}_\pi(E')=\frac{(m^2_\tau+m^2_\pi)(|\vec k\,|+M_n-E')\pm
(m^2_\tau-m^2_\pi)\sqrt{(|\vec k\,|+M_n-E')^2-m^2_\tau}}{2m^2_\tau}\label{eq:Epionlimits} 
\eea
Besides,  the $H(1-|\cos\varphi^0_\pi|)$ function restricts
$\cos\theta_\pi$  to the interval $\cos\theta_\pi\in[\cos(\theta^0_{\tau\pi}+\theta^0_\tau),
\cos(\theta^0_{\tau\pi}-\theta^0_\tau)]$. In addition we have
\bea
\sin\theta^0_{\tau}\sin\theta_\pi|\sin\varphi^0_\pi|&=&
\sqrt{1-\cos^2\theta_\pi-\cos^2\theta^0_{\tau}-\cos^2\theta^0_{\tau\pi}+2
\cos\theta_\pi\cos\theta^0_{\tau}\cos\theta^0_{\tau\pi}}\no
&=&\sqrt{(\cos\theta_\pi-\cos(\theta^0_{\tau\pi}+\theta^0_\tau))
(\cos(\theta^0_{\tau\pi}-\theta^0_\tau)-\cos\theta_\pi)}\ .
\eea
Taking all these considerations into account, we arrive at the final expressions given in Eqs.~\eqref{eq:sig1}–\eqref{eq:sig2} for the triple-differential cross section  $d^3\sigma/(dE'dE_\pi d\cos\theta_\pi)$, and, analogously, for the asymmetry $\Delta\sigma$. 

One can increase the statistics of the pion energy and angular distributions by performing first the $dE'$ integral in Eqs.~\eqref{eq:sig1}-\eqref{eq:sig2}. To this end, we interchange the order of integration,
\bea
\sigma&=&{\cal B}(\tau\to\pi\nu_\tau)\frac{ m^2_\tau M_{\Lambda_c}G_F^2|V_{cd}|^2}
{2\pi^2|\vec k\,|^2(m^2_\tau-m^2_\pi)}
\int_{E^{\rm min}_\pi}^{E^{\rm max}_\pi}dE_\pi\int_{-1}^1d\cos\theta_\pi\ 
\frac{\Theta(\cos\theta_\pi,E_\pi)}{\sqrt{|A|}}
\int_{E^{\prime}_1}^{E^{\prime}_2}dE'
\\
&\times&\frac{\sum_{\chi=L,R}\rho_{\chi\chi}\sum_{a,b}W^{ab}_\chi L_{\chi ab}}
{\sqrt{(E'-E^{\prime}_1)(E^{\prime}_2-E')}}\Big(1
+\frac{2m_\tau}{m^2_\tau-m^2_\pi}\Big\{\frac{{\cal P}_L }{m_\tau}
[E_\pi
\sqrt{(|\vec k\,|+M_n-E')^2-m^2_\tau}
\nonumber \\
&-&
|\vec p_\pi|(|\vec k\,|+M_n-E')
\cos\theta^0_{\tau\pi}]+ \frac{{\cal P}_T|\vec p_\pi|}{\sin\theta^0_\tau}(\cos\theta_\pi-
\cos\theta^0_\tau\cos\theta^0_{\tau\pi})\Big\}\Big),\label{eq:sec3}\\
\Delta\sigma&=&{\cal B}(\tau\to\pi\nu_\tau)\frac{ m^2_\tau M_{\Lambda_c}G_F^2|V_{cd}|^2}
{2\pi^2|\vec k\,|^2(m^2_\tau-m^2_\pi)}
\int_{E^{\rm min}_\pi}^{E^{\rm max}_\pi}dE_\pi\int_{-1}^1d\cos\theta_\pi
\frac{\Theta(\cos\theta_\pi,E_\pi)}{\sqrt{|A|}}
\int_{E^{\prime}_1}^{E^{\prime}_2}dE'
\nonumber\\
&\times&\frac{\sum_{\chi=L,R}\rho_{\chi\chi}\sum_{a,b}W^{ab}_\chi L_{\chi ab}}
{\sqrt{(E'-E^{\prime}_1)(E^{\prime}_2-E')}}
\,\frac{2m_\tau}{m^2_\tau-m^2_\pi}\Big[-{\cal P}_{TT}\, |\vec p_\pi|\,
|\sin\varphi^0_\pi|\,\sin\theta_\pi\Big],
\label{eq:asymmetry}
\eea
where we have written
\bea
(\cos\theta_\pi-\cos(\theta^0_{\tau\pi}+\theta^0_\tau))
(\cos(\theta^0_{\tau\pi}-\theta^0_\tau)-\cos\theta_\pi)&=&
1-\cos^2\theta_\pi-\cos^2\theta^0_\tau-\cos^2\theta^0_{\tau\pi}+2
\cos\theta_\pi\,\cos\theta^0_\tau\,\cos\theta^0_{\tau\pi}\no
&=&
\frac{AE^{\prime 2}+BE'+C}{(|\vec k\,|+M_n-E')^2-m^2_\tau}=
\frac{|A|(E'-E^{\prime }_1)(E^{\prime }_2-E')}{(|\vec
k\,|+M_n-E')^2-m^2_\tau},
\eea
with the numerator being a second-order polynomial $AE^{\prime 2}+BE'+C$ in the  $\Lambda_c$ 
energy, with $E^{\prime}_2>E^{\prime}_1$ its two roots, and 
where we have taken into account 
that the $A$ coefficient above, which is given by
\be
A=\sin^2\theta_\pi-\Big(\frac{|\vec k\,|+M_n}{|\vec k\,|}\Big)^2-\frac{E^2_\pi}{|\vec p_\pi|^2}+
2\cos\theta_\pi\frac{E_\pi(|\vec k\,|+M_n)}{|\vec k\,||\vec p_\pi|},
\ee
 is negative.\footnote{Note its maximun is reached for $\theta_\pi=0$, being its value at that maximum
 \be
 -\Big(\frac{|\vec k\,|+M_n}{|\vec k\,|}-\frac{E_\pi}{|\vec p_\pi|}\Big)^2\le0,
 \ee 
 with the equality holding only for $\cos\theta_\pi=1$ and $E_\pi=m_\pi(|\vec k\,|+M_n)/\sqrt{M^2_n+
 2M_n|\vec k\,|}$ which is a null set.}
The $\Theta(E_\pi,\cos\theta_\pi)$ function is 1  if $B^2-4AC\ge0$ and zero
otherwise, guaranteeing in this way that $E'_{1,2}$ are real. 
Besides, one has\footnote{For the pion to be at rest,  the
energy of the intermediate $\tau$ has to be $(m^2_\tau+m^2_\pi)/(2m_\pi)$. However,
for this to happen one needs that $|\vec k\,|\ge \hat k_0$. The $\hat k_0$ 
value derives from
the forward scattering equality $\hat k_0+M_n=(m^2_\tau+m^2_\pi)/(2m_\pi)+
\sqrt{M^2_{\Lambda_c}+(\hat k_0-(m^2_\tau-m^2_\pi)/(2m_\pi))^2}$.}
\be
E^{\rm max}_\pi=E^+_\pi(E'_-),\ 
E^{\rm min}_\pi=\left\{\begin{array}{ll}E^{-}_\pi(E'_-) &
 |\vec k\,|\le\hat k_0\\
m_\pi&|\vec k\,|>\hat k_0\end{array}\right.
,\
\hat k_0=\frac{m^2_\tau(M_n-m_\pi)+M_nm^2_\pi+m_\pi(M^{2}_{\Lambda_c}-M^2_n)}
{2(M_n-m_\pi)m_\pi}.
\label{eq:hatk0}
\ee

\section{ Correspondence between the $\bar\nu_\tau p\to\Lambda \tau^+(\pi^+\bar\nu_\tau)$ and $\nu_\tau n \to \Lambda_c \tau^- (\pi^- \nu_\tau)$ cross sections.}
\label{app:anti}
In this case we have for the $\tau^+\to \pi^+\bar\nu_\tau$ decay the effective vertex
\bea
{\cal L}_{\bar\tau\to\pi\bar\nu_\tau} = i2G_Ff_\pi V_{ud}\bar\Psi_\tau(x)\gamma^\mu
P_L\Psi_{\nu_\tau}(x)\partial_\mu\Phi^+(x)=
-i2G_Ff_\pi V_{ud}\overline{\Psi^{\cal C}}_{\nu_\tau}(x)\gamma^\mu
P_R\Psi^{\cal C}_{\tau}(x)\partial_\mu\Phi^+(x),
\eea
where the last expression is written in terms of the $\tau$ and $\nu_\tau$ 
charge-conjugated fields.
With respect to the reaction analyzed in Sect.~\ref{sec:Lambdac}, and apart from the obvious
changes in the values of the CKM matrix elements and  masses, form factors, Wilson coefficients and density-matrix 
matrix elements that enter the evaluation of $\sum_{\chi=L,R}\rho_{\chi\chi}\sum_{a,b}W^{ab}_\chi 
L_{\chi ab}$, we now have the extra modification
\bea
&&{\rm Tr}\,[\slashed{p}_{\nu_\tau} P_R(\slashed{k}'+m_\tau)
\Gamma_a \frac1{2}P_\chi\slashed{k}\gamma^0\Gamma^\dagger_b\gamma^0
(\slashed{k}'+m_\tau)P_L]\longrightarrow {\rm Tr}\,[\slashed{p}_{\nu_\tau} P_L(\slashed{k}'+m_\tau)
\Gamma_a \frac1{2}P_\chi\slashed{k}\gamma^0\Gamma^\dagger_b\gamma^0
(\slashed{k}'+m_\tau)P_R]
\label{eq:trazaanti}
\eea
with $p_{\nu_\tau}, k'$ representing the four momenta of the final
$\bar\nu_\tau$ and intermediate $\tau^+$ respectively. We get 
 for that trace $(p_{\nu_\tau}=k'-p_\pi)$
\bea
\frac{m_\tau^2-m_\pi^2}2 {\rm Tr}\,[(\slashed{k}'+m_\tau)
\Gamma_a \frac12P_\chi\slashed{k}\gamma^0\Gamma^\dagger_b\gamma^0]
-\frac12{\rm Tr}\,[(\slashed{k}'+m_\tau)
\Gamma_a \frac1{2}P_\chi\slashed{k}\gamma^0\Gamma^\dagger_b\gamma^0
(\slashed{k}'+m_\tau)\gamma_5\slashed{p}_\pi].
\eea
Comparison with Eq.~\eqref{eq:appearP} thus leads to the additional modification specified in Eq.~(\ref{eq:change}).

We note that Eq.~(\ref{eq:change}) is consistent with the results of Eq.~(10) in Ref.~\cite{Hernandez:2022nmp}, where the similar
$\nu_\tau A_Z\to\tau^-(\pi^-\nu_\tau)X$ and $\bar\nu_\tau A_Z\to\tau^+(\pi^+\bar\nu_\tau)
X$ reactions, driven by the $d\to u$ and $u\to d$ transitions respectively, were 
analyzed within the SM. It might look like that,  for the case of the antineutrino 
induced $u\to d$ reaction, the  expression of Ref.~\cite{Hernandez:2022nmp}  and that 
found here differ in an overall sign in the $p_\pi\cdot{\cal P}$ contribution.  However, 
both calculations are consistent, and the apparent  sign discrepancy disappears when ones
takes into account the fact that  for the SM contribution, the only one considered in 
Ref.~\cite{Hernandez:2022nmp}, the $P_\chi$ projector in our present 
Eq.~(\ref{eq:trazaanti}) is  $P_R$. This is because, within the SM, all combination 
of Wilson coefficients in  Eq.~\eqref{eq:wcsutau} are zero, except for the right-handed 
$C^V_{su\tau R}$ and $C^A_{su\tau R}$, which take the values $-1$ and 1, respectively. 
In contrast, the SM contribution to the polarization vector is purely left-handed for the 
neutrino induced $d\to u$ transition, since in  that case,  all combination of Wilson
coefficients are zero, except for $C^V_{L}$ and $C^A_{L}$ 
which are both equal to one (see the similar case in our Eq.~\eqref{eq:wcdctau}). Next, 
one should take into account that right and 
left-handed contributions to ${\cal A}_{{\cal H}},{\cal B}_{{\cal
H}}, {\cal C}_{{\cal H}}$ and ${\cal D}_{{\cal H}}$ functions, which appear in the 
calculation of the SM ${\cal P}^\mu$, differ in an overall sign, as can be seen in 
Eq.~(D.2) of Ref.~\cite{Penalva:2021wye}, and it compensates the overall sign in 
the $p_\pi\cdot{\cal P}$ term of  Eq.~\eqref{eq:change}.\footnote{This is 
because of the factor $h_\chi$ in the $\sum_{\chi=L,R}$ of Eq.~(D.2) of 
Ref.~\cite{Penalva:2021wye}. In addition, we can see in Eq.~(D.1) of the same 
reference, that this distinctive sign ($h_\chi$) does not appear  in the sum of 
left and right-handed contributions  to ${\cal A}$, ${\cal B}$ and ${\cal C}$, which
 determine  the overall $\sum_{\chi=L,R}\rho_{\chi\chi}\sum_{a,b}W^{ab}_\chi 
L_{\chi ab}$ factor. Finally, we note that when only SM is 
 considered, we have $C^V_{L}C^A_{L}=1$ for neutrinos, while $C^V_{R} C^A_{R}=-1$ 
 for antineutrinos. This is responsible for the $\pm$ differences in  
 ${\cal P}^{\tau,\bar\tau}$ seen in Eqs.~(7) and (8) of Ref.~\cite{Hernandez:2022nmp}.}
\bibliography{LCTAU}

\end{document}